\documentclass[11pt,a4paper]{article}
\usepackage[T1]{fontenc}
\usepackage[utf8]{inputenc}
\usepackage{authblk}

\usepackage{amssymb}
\usepackage{algorithm}
\usepackage{algorithmic}
\usepackage{clrscode}
\usepackage{amsmath}
\usepackage{multirow}
\usepackage{courier}
\usepackage{graphicx}

\usepackage{geometry}

\usepackage{url}
\providecommand{\keywords}[1]
{
  \small	
  \textbf{\textit{Keywords---}} #1
}

\begin{document}

\title{A Survey on Contact Tracing: the Latest Advancements and Challenges}

% \date{}
% \author{\uppercase{Ting Jiang}\authorrefmark{1},\uppercase{Yizhen Chen}\authorrefmark{1}, \uppercase{Yang Zhang}\authorrefmark{1},\uppercase{Minhao Zhang}\authorrefmark{1},\uppercase{Chenhao Lu}\authorrefmark{1},\uppercase{Ji Zhang}\authorrefmark{2},\uppercase{Zhao Li}\authorrefmark{3},\uppercase{Jun Gao}\authorrefmark{4},\uppercase{Shuigeng Zhou}\authorrefmark{5}}
% \address[1]{Zhejiang Lab, Hangzhou, China}
% \address[2]{The University of Southern Queensland, Toowoomba, Australia}
% \address[3]{Alibaba Group, Hangzhou, China}
% \address[4]{Peking University, Beijing, China}
% \address[5]{Fudan University, Beijing, China}
% \markboth
% {Mingming Lu\headeretal: Preparation of Papers for IEEE TRANSACTIONS and JOURNALS}
% {Mingming Lu \headeretal: Preparation of Papers for IEEE TRANSACTIONS and JOURNALS}

% \corresp{Corresponding author: Ji Zhang (e-mail: zhangji77@gmail.com).}

% \author{Ting Jiang \and Yizheng Chen \and Yang Zhang \and Minhao Zhang \and Chenhao Lu \and Ji Zhang
% \thanks{Corresponding author.} \and Zhao Li \and Jun Gao \and Shuigeng Zhou}

\author[1]{Ting Jiang}
\author[1]{Yang Zhang}
\author[1]{Minhao Zhang}
\author[1]{Ting Yu}
\author[1]{Yizheng Chen}
\author[1]{Chenhao Lu}
\author[2]{Ji Zhang \thanks{Corresponding author: zhangji77@gmail.com}}
\author[3]{Zhao Li}
\author[4]{Jun Gao}
\author[5]{Shuigeng Zhou}
% \affil[a]{Department of Computer Science, \LaTeX\ University}
% \affil[b]{Department of Mechanical Engineering, \LaTeX\ University}
\affil[1]{Zhejiang Lab, Hangzhou, China}
\affil[2]{The University of Southern Queensland, Toowoomba, Australia}
\affil[3]{Alibaba Group, Hangzhou, China}
\affil[4]{Peking University, Beijing, China}
\affil[5]{Fudan University, Beijing, China}

\renewcommand*{\Affilfont}{\small\it} % 修改机构名称的字体与大小
\renewcommand\Authands{ and } % 去掉 and 前的逗号
\date{} % 去掉日期
% \author{Ting Jiang\inst{1} \and Yizheng Chen\inst{1} \and Yang Zhang\inst{1} \and Minhao Zhang\inst{1} \and Chenhao Lu\inst{1} \and Ji Zhang\inst{2}
% \thanks{Corresponding author.} \and Zhao Li\inst{3} \and Jun Gao\inst{4} \and Shuigeng Zhou\inst{5}}

% \institute{Zhejiang Lab, Hangzhou, China
% \and The University of Southern Queensland, Toowoomba, Australia
% \and Alibaba Group, Hangzhou, China \andPeking University, Beijing, China \and Fudan University, Beijing, China}
% \author[label1]{Ting Yu}
% \ead{yuting@zhejianglab.com}

% \author[label2]{Mengchi Liu\corref{cor1}}
% \ead{mengchi@scs.carleton.ca}

% \author[label1]{Ji Zhang\corref{cor1}}
% \ead{Ji.Zhang@zhejianglab.com}

% \author[label1]{Zujie Ren}
% \ead{zujie.rzj@zhejianglab.com}

% \cortext[cor1]{Corresponding author.}
% \address[label1]{Zhejiang Lab, Hangzhou, PRC}
% \address[label2]{School of Computer Science, Carleton University, 1125 Colonel By Drive, Ottawa Ontario K1S 5B6, Canada}

% \renewcommand{\shortauthors}{Ting Jiang and Yizhen Chen, et al.}
\maketitle
\begin{abstract}
Infectious diseases are caused by pathogenic microorganisms, such as bacteria, viruses, parasites or fungi, which can be spread, directly or indirectly, from one person to another. Infectious diseases pose a serious threat to human health, especially COVID-19 that has became a serious worldwide health concern since the end of 2019. Contact tracing is the process of identifying, assessing, and managing people who have been exposed to a disease to prevent its onward transmission. Contact tracing can help us better understand the transmission link of the virus, whereby better interrupting its transmission. Given the worldwide pandemic of COVID-19, contact tracing has become one of the most critical measures to effectively curb the spread of the virus. This paper presents a comprehensive survey on contact tracing, with a detailed coverage of the recent advancements the models, digital technologies, protocols and issues involved in contact tracing. The current challenges as well as future directions of contact tracing technologies are also presented.
\end{abstract}


\section{Introduction}
\begin{figure}
  \centering
  \includegraphics[width=13cm]{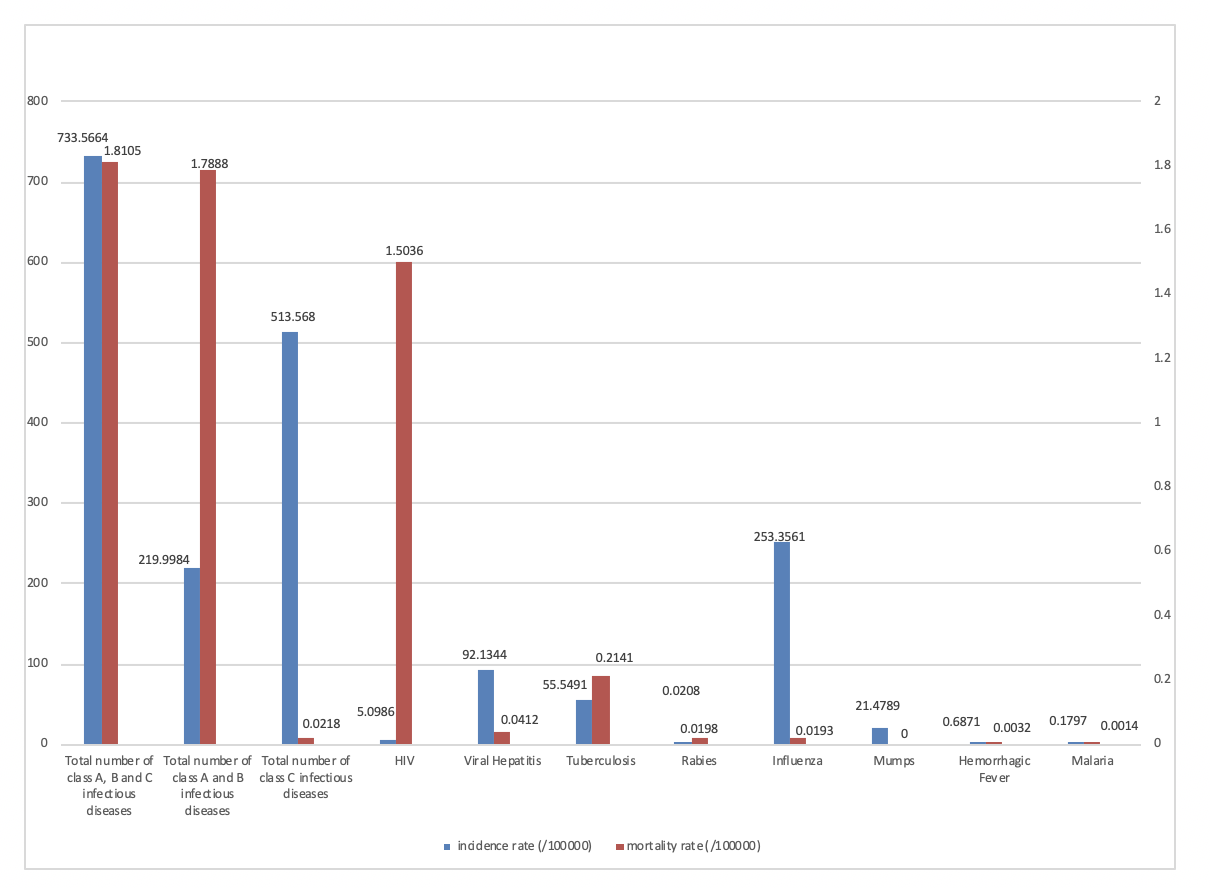}
  \caption{Statistics on the incidence and death of China's statutory infectious diseases in 2019.}
  \label{fig:1}
\end{figure}

Infectious diseases are an illness resulting from pathogenic microorganisms, also known as transmissible or communicable diseases. The pathogens that cause infectious diseases are bacteria, viruses, parasites, fungi and so on \cite{infectious}. Infectious diseases bring a great threat to human health, national economy and societal development. Death from infectious diseases is one of the top 10 causes of death worldwide, and the incidence of infectious diseases in developing countries is especially high \cite{threat}. In 2016, infectious diseases resulted in 4.3 million deaths (1.7 million women and 2.7 million men) according to the WHO's 2019 World Health Statistics report \cite{who1}. According to the Statistics of the National Health Commission of China in 2019, a total of 1,0244,507 cases of statutory infectious diseases were reported nationwide, resulting in 25,285 deaths. The reported morbidity rate was 733.57 per 100,000 and the mortality rate was 18,100 per 100,000. 
%%%comment 3.1
As shown in Figure \ref{fig:1}, HIV Viral Hepatitis Tuberculosis Rabies influenza are among the top five infectious disease deaths in China. Starting from the end of 2019, COVID-19 has taken a huge death toll on people's life, having caused 45,406,832 infections and 1,183,726 deaths as of October 30, 2020 \cite{Livebroadcast}. 
%Contact tracing can better interrupt the transmission of the virus \cite{classoftran}. Contact tracing has been used in pandemics such as HIV/AIDS, Ebola , tuberculosis and etc. Therefore, contact tracing of infectious disease investigation is of great significance. %%add referent
%%%comment 1.5+3.1
Infectious diseases can be categorized in different ways. Here, we present two commonly used approaches for categorizing infectious diseases. According to the mode of transmission, infectious diseases can be broadly divided into the following four categories \cite{classoftran}:
\begin{itemize}
    \item \textbf{Respiratory tract infectious diseases,} such as SARS, measles and influenza, which can be transmitted through air, droplets and contact with respiratory secretions, etc.; 
    \item \textbf{Digestive infectious diseases,} such as paratyphoid A and B, foodborne diseases and esherihioz, which can be transmitted through mosquitoes, flies or fleas;
    \item \textbf{Transmissive or blood infectious diseases,} such as malaria, plague and HIV infection, which can be transmitted through blood;
    \item \textbf{Surface infectious diseases,} such as tetanus, anthrax and infection with multiple pathways like infectious mononucleosis and enterovirus infections, which can be transmitted through the direct contact with the pathogen.
\end{itemize}

According to the diseases' contagiousness, spread and risk of death, China classifies infectious diseases into three main categories, called Class A, B and C infectious diseases as following \cite{TypesofDiseases}:
\begin{itemize}
    \item \textbf{Class A infectious diseases,}which are subject to compulsory management, such as plague and cholera;
    \item \textbf{Class B infectious diseases,}which are subject to stringent management, such as viral hepatitis, typhoid and paratyphoid, AIDS, gonorrhea, syphilis, etc;
    \item \textbf{Class C infectious diseases,} which are subject to close monitoring, such as tuberculosis, leprosy, influenza, epidemic parotitis and neonatal tetanus.
\end{itemize}

Contact tracing is the process of identifying, assessing, and managing people who have been exposed to an infectious disease to prevent onward transmission. When systematically applied, contact tracing will break the chains of transmission of an infectious disease and is thus an essential public health tool for controlling infectious disease outbreaks \cite{who2}. It is a critical process to ensure the best possible chance of control and the longest possible time to local take-off \cite{muller2000contact}. The importance of contact tracing, timely testing and adequate quarantine/isolation in disease control has been proven \cite{2020Contact}. The WHO had defined contact tracing as having three basic steps:
\begin{itemize}
    \item \textbf{Contact identification, } where the infected person recalls activities and the roles of persons involved since the onset of the infectious disease. Contacts can be anyone who has been in close contact with an infected person, including, but not limited to, the family members, work colleagues, friends or health care providers;
    \item \textbf{Contact listing,} which provides the names of potentially infected contacts;
    \item \textbf{Contact follow-up,} which monitors any onset of symptoms associated with the viral infection \cite{who3}.
\end{itemize}

%%%comment 3.2
In history, contact tracing has been widely used in the control of infectious diseases and has become a pillar of communicable disease control in public health for decades. An HIV-positive obstetrician in Australia infected 149 pregnant women through contact in 1994. This event attracted wide attentions, and helped prompt more research attentions on controlling infectious diseases through contact tracing \cite{Julianne1996Infinitesimal}. Contact tracing and the follow-up control measures such as quarantine and isolation were crucially important during the SARS outbreak in 2003 \cite{riley2003transmission}, the Ebola outbreak in Africa in 2014 \cite{saurabh2017role}, as well as in the eradication of smallpox \cite{2003Contact}. Despite the current advances in vaccine development technologies, the role of contact tracing and follow-up control measures in the initial stage of an epidemic are still critical. The results show that a response system based on enhanced testing and contact tracing can have a major role in relaxing social-distancing interventions in the absence of herd immunity against SARS-CoV-2 \cite{2020Modelling}.

Nevertheless, the traditional contact tracing approaches used to control the spread of infectious disease requires a lot of human and material resources, which is disadvantageous given that the fast response is critical to deal with those highly infectious diseases. The more recent digital contact tracing methods are far more efficient and have helped greatly reduce the consumption of human labor and other material resources. Given their great advantages, various digital information technologies in contact tracing have been applied in many countries as
%%%comment 3.3
effective means for COVID-19 inhibition \cite{who5}. 
%The contact tracing method used in different countries has different advantages and disadvantages. It will be helpful to novel Coronavirus control if we can study the effectiveness of contact tracing and the methods of using contact tracing in different countries.
%As a result, some researchers have proposed using a mathematical model for contact-tracking, which does not require a lot of resources.
          
To our best knowledge, the previous work on surveying contact tracing, particularly for COVID-19, is rather limited. The only major survey work that we can find in our literature review is the survey work carried out by Nadeem et al. \cite{ahmed2020survey}. Yet, this survey is very limited in its scope and depth - it only covered some of the COVID-19 contact tracing apps and protocols. In comparison, our survey is much more comprehensive, covering both the traditional and the recent digital contact tracing technologies. Our survey presents the fundamental approach and models of contact tracing and the latest digital contact tracing technologies, detailing a variety of tracing systems and apps currently being used in many different countries, together with the associated localization technologies. The important issues and protocols involved in digital contact tracing are also discussed. At the end of the survey, we reflect on the current challenges contact tracing technologies are facing and highlight several future trends projected for technological developments of contact tracing in the future.

\textbf{Structure of our survey.} The rest of our survey is organized as follows.
In Section 2, we introduce the traditional contact tracing method and some underlying theoretical models that the traditional contact tracing methods have used.
%%%comment 3.6
The important privacy issues and protocols in contact tracing are discussed in Section 3. In Section 4, we elaborate on a variety of the latest contact tracing apps currently being used in many different countries as well as the associated localization technologies. Section 5 covers some other important issues involved in contact tracing technologies. We identify the current challenges and future directions of contact tracing technologies in Section 6. Finally, Section 7 concludes this survey.

\section{Traditional Contact Tracing Methods, Technologies and Models}
%%(引言部分）阐述contact tracing使用的传统tech以及理论的东西

In this section, we will present some preliminaries of contact tracing, including the traditional contact tracing methods and the underlying mathematical models, both of which have been traditionally utilized by medical and healthcare professionals for carrying out contact tracing. 
%%%comment 3.4
\subsection{Traditional Contact Tracing Methods}
\par The fundamental principle in contact tracing is nothing more than the old-school detective work - finding sick patients and then figuring out who they recently interacted with, even though the exact approach implemented in contact tracing may vary.
 
\par The most popular and well-known contact tracing method is the backward tracing. It seeks to establish how someone became infected in the first place, rather than the persons to whom the infection has been passed to. For epidemics with high heterogeneity in infectiousness, experts sometimes may adopt a hybrid strategy combining both the forward and backwards tracing to find the source of infection, depending on local contact tracing capacity. A typical backward contact tracing strategy is based on a community-based survey and follows a certain steps as follows:
    \begin{itemize}
        \item When an individual is identified as being infected or as a virus carrier, he/she will be reported to the relevant public health organisation and be put under quarantine;
        \item The individual will be interviewed by contact tracers to establish the history of his movements and the close contact whom might be infected;
        \item Contact tracers may also need to interview those who have information about the patients’ contact. Considering the possible cases of the interviewee's intentional concealment or his inability of recalling the details, necessary cross validation will be carried out;
        \item Once contacts are identified, contact tracers will contact them and test them for possible infection;
        \item Once the contact is also identified as being infected by the disease, then contact tracers will start again from the first step to carry on the tracing;
        \item If contacts cannot be individually identified, then some broader communications may be issued to the public within a certain proper scope depends on how widely the virus has possibly spread.
    \end{itemize}
    
In the tracing process, case management software is often used by contact tracers to maintain records of cases and contact tracing activities, which are typically stored in a centralized or cloud-based database to support fast information browsing and retrieval. 

As far as the contact notification is concerned, it includes patient referral and provider referral in the traditional contact tracing routine, according to the Australasian contact tracing guidelines \cite{mh201}. Table \ref{contacts notification} provides a more detailed explanation about this. The contact tracing management software may also have some special features that can use SMS or email directly to notify people believed to have been in close contact with someone who has been contracted by the infectious disease.

%     \par \textbf{Patient referral}
% 	    \begin{itemize}
% 	    \item The index patient personally notifies his or her contact.
% 	    \item The health care provider provides the information to be imparted by the index patient to the partner.
% 	    \end{itemize}
%     \par Advantages
% 	    \begin{itemize}
% 	    \item   Individuals usually prefer to notify contacts personally.
% 	    \item   Quicker and easier.
% 	    \end{itemize}
%     \par Disadvantages
% 	    \begin{itemize}
% 	    \item   Less confidentiality
% 	    \item   Patients may not actually contact partners
% 	    \end{itemize}
%     \par \textbf{Provider referral}
% 	    \begin{itemize}
% 	    \item   The health care provider directly advises the contact or uses another agency (for example, sexual health service, public health unit or health department contact tracer) to ensure that contacts are notified.
% 	    \item   The health care provider must have the explicit approval of the index patient.
% 	    \end{itemize}
%     \par Advantages
% 	    \begin{itemize}
% 	    \item    Higher level of confidentiality for the index patient.
% 	    \item    Method of choice when an individual fears a violent reaction, and for certain situations and conditions (for example, pulmonary TB, transfusion-related infections, when contact will involve sex workers or person with intellectual disability).
% 	    \item   May be appropriate for serious infections such as HIV where rigorous case finding is warranted.
% 	    \end{itemize}
%     \par Disadvantages
%     \begin{itemize}
%     \item   More time- and resource-intensive.
% 	\end{itemize}

	\begin{table}
      \centering
      \caption{contacts notification.}
      \label{contacts notification}
      \begin{tabular}{p{1.5cm}p{5cm}p{6cm}}
        \hline
          & \textbf{Patient referral} & \textbf{Provider referral}\\
        \hline
        \textbf{Process} & \begin{itemize}
	    \item The index patient personally notifies his/her contact.
	    \item The healthcare provider provides the information to be imparted by the index patient to the partner.
	    \end{itemize} &\begin{itemize}
	    \item   The healthcare provider directly advises the contact or uses another agency (e.g., sexual health service, public health unit or health department contact tracer) to ensure that contacts are notified.
	    \item   The healthcare provider must have the explicit approval of the index patient.
	    \end{itemize} \\
        \textbf{Advantages} &\begin{itemize}
	    \item   Individuals usually prefer to notify contacts personally.
	    \item   Quicker and easier.
	    \end{itemize} & \begin{itemize}
	    \item    Higher level of confidentiality for the index patient.
	    \item    Method of choice when an individual fears a violent reaction, and for certain situations and conditions (e.g., pulmonary TB, transfusion-related infections, when contact will involve sex workers or person with intellectual disability).
	    \item   May be appropriate for serious infections such as HIV where rigorous case finding is warranted.\end{itemize}\\
        \textbf{Disadvantages} &\begin{itemize}
	    \item   Less confidentiality.
	    \item   Patients may not actually contact partners.
	    \end{itemize} & \begin{itemize}
    \item   More time and resource intensive.
	\end{itemize}\\
      \hline
    \end{tabular}
    \end{table}     
	
The above fundamental contact tracing method has been widely performed on contact tracing for diseases such as tuberculosis, vaccine-preventable infections (e.g., measles), sexually transmitted infections (e.g., HIV), blood-borne infections, Ebola, some serious bacterial infections, and novel virus infections (e.g., SARS, H1N1, Covid-19). 
	   
    %\par The traditional methods are always closely related to the circumstance and environment according to K.O. Kwok et al., which means there are limited circumstances we can implement \cite{mh114}.
    
    Even though it has been proven effective in handling many infectious diseases in the past, the traditional contact tracing method suffers from some major limitations when dealing with large-scale outbreak of infectious diseases such as the pandemic of COVID-19. Relying mainly on manual interviews and investigations, the traditional tracing method is time and resource consuming. A low, steady number of new cases is usually manageable by the traditional contact tracing methods, but for those highly infectious diseases, such as COVID-19, the infection rate of the diseases can easily outpace that of the contact tracing and will quickly overwhelm the tracing system \cite{ch04}, making contact tracing lose much of its efficacy eventually. Also, the traditional contact tracing method is mainly based on interviews and thus tends to produce highly incomplete and inaccurate information. This will mislead or even disrupt the tracing process and waste a lot of resources unnecessarily which may have already stretched to the limit.
    
    Therefore, it becomes imperative for us to develop and deploy more efficient contact tracing technologies, by taking advantage of the latest advancements in information and communication technologies, for faster responses to major infectious disease outbreak. 

    %%%comment 1.1+3.5
    \subsection{Underlying Fundamental Models For Contact Tracing}
    
    As for its theoretical foundation, contact tracing often draws on some underlying models that reveal the dynamics of virus transmission. For example, mathematical models were developed to study the dynamics of SARS and MERS transmissions in early 2000, which provide important support to contact tracing for the two diseases. Models represented how contact tracing affected the epidemic dynamics were useful for evaluating different infection control interventions and the burden of infection to facilitate the further understanding of their epidemiology \cite{mh114}. Such models have direct utility in planning for future outbreaks of coronaviruses: they can be used to estimate the scale of resources required to conduct effective contact tracing.
    
    %\subsubsection{Networks Models.} 
    
    %\par Traditional methods are fundamentally linked to the individual-level spread of infection and, in particular, the network of potential transmission routes.
    
     %\par In homogeneous networks, when all infectious individuals display symptoms and all individuals have an exactly contacts number to whom they can pass infection. By looking at the parameter values at which equilibrium prevalence and the initial growth of infection are zero, experts can derive the level of tracing required to eliminate infection.
     
    %\subsubsection{Mathematical Models}
    Since 1991, Kermack et al. \cite{mh7} have demonstrated that mathematical theory has made a significant contribution to epidemic control. Many mathematical models of epidemic have sprung up. Currently, a large part of mathematical epidemiology is concerned with the investigation of mechanisms and efficacy of control strategies against infectious diseases. They have proved themselves as powerful tools, helping human beings to control the epidemic situation, but many of them are found to have problems of low efficiency and high cost.
        
    In 2021, Johannes Müller and Mirjam Kretzschmar \cite{mh1.1} discussed how they sort contact tracing models based on the first principles, which means models derived from an individual-based stochastic model where contact tracing can be directly formulated.
    % but rather define: A phenomenological term is introduced, and without deeper justification it is claimed that this term describes contact tracing adequately.

    Three main directions of contact tracing models can be identified based on the first principles:
        \begin{itemize}
            \item Individual based simulation models that directly simulate individuals in a large populations;
            \item Pair approximation models that connect the correlations of pairs of individuals for incorporating information;
            \item Stochastic and deterministic models that are based on branching process modelling.
        \end{itemize}

        %Specifically, the contact tracing models for epidemic control are mainly divided into stochastic models, deterministic models and other models. Castillo-Chavezet et al. have written a book which covers the use of dynamical systems (deterministic discrete, delay, PDEs, and ODEs models) and stochastic models in disease dynamics \cite{mh8}.
            \paragraph{a}Individual based models
            
            Individual based models are most basic yet popular to formulate a process as complex as contact tracing. They trace down every movement of individuals and their interactions. In doing so, they incorporate a graph where individuals form the nodes, connected by edges that describe the possible contacts between them. For each edge, a stochastic process indicates the time points of contacts. The contact graph can be as simple as a complete graph where every individual may have contact with every other individual, a random graph as described by the configuration model, or a small world graph that reflects local and long distance contacts \cite{mh1.2} \cite{mh1.3}.
            
            As a general and popular data structure to represent complex relationships between entities such as in society \cite{jz03}\cite{jz04} and biology \cite{jz05}, network has been also an important concept in individual based models, where social interactions can take place over a wide range of distances. While some models have been developed to consider the role of contact tracing in randomly interacting populations \cite{muller2000contact}, only network-based models that consider transmission pathways \cite{mh116} and the associated pairwise equations can provide an accurate mechanistic understanding of the structured nature of human interactions.
            
            In one-step tracing, only the nearest neighbour is considered in the tracing process. Meanwhile, recursive will trace an identified infected neighbour when it becomes a new index case. In such constructed models, there is almost no limit to the degree of detail that can be included. However, too much details often leads to the problem of lack of data for an appropriate parametrization.
            
            \paragraph{b}Pair approximation models
            
            Pair approximation is a method to derive ordinary differential equations that describe stochastic dynamics by allowing the considering both the frequency of individuals as well as the frequency of pairs of individuals. These equations are a well established, heuristic approach to reformulate individual based models as described above. Instead of counting the number of individuals of a given type, the expected relative frequencies are addressed by the ordinary differential equations model. In many cases, these models allow for a deeper understanding of the underlying mechanism, and for powerful predictions \cite{mh1.4}. Detailed pairwise equations \cite{mh117} give an analytical insight into contact tracing for STDs, such as gonorrhoea and chlamydia, which are described by a susceptible–infected–susceptible (SIS) framework, while the airborne infections, such as SARS, smallpox, polio and measles, follow the susceptible–infected–recovered/dead paradigm.
            
            \paragraph{c}Stochastic Models
            
            Stochastic models, as well as the Markov Chain Monte Carlo (MCMC) methods, have become increasingly popular for calibrating stochastic epidemiological models with missing data which treat the missing data as extra parameters \cite{mh12}\cite{mh13}\cite{mh14}. Istvan Z. Kiss et al. thought that the existing stochastic model does not consider the transmission of between groups, and put forward a stochastic model with better clustering effect for epidemic diseases transmitted by groups \cite{mh17}. For stochastic models, super spreading events were found pretty difficult to be considered in the models. 
             
	        \paragraph{d}Deterministic Models
            
            Yorke et al. constructed a deterministic model for the spread of gonorrhea in a community by considering only the sexually active individuals who may potentially contract the disease from their contacts \cite{mh10}. Although deterministic modeling can be a guide for describing epidemics, parameter estimation for deterministic models is usually a difficult task because of missing observations.
            
            For deterministic models like discrete-time simulation models, there exists overestimation in contact tracing efficacy due to failure of identifying correlation structure between diseases generation by contact tracing.
    \section{Privacy issues and protocols}
    %%%增加引导语言
    Some respiratory infectious diseases, such as COVID-19, are highly infectious and can spread very quickly if not properly contained. Even worse, someone who infected by the virus can go unnoticed for up to 20 days before showing symptoms or being tested positive, before when the infected individual may have already contacted many other people and passed the disease on to them. Therefore, it is of a paramount importance to quickly and accurately to carry out contact tracing to understand how the virus is transmitted and notify the contacted individuals as soon as possible.
  
    Digital contact tracing is a contact tracing method that relies on mobile devices to determine the contact between infected patients and users, which uses some contact tracing technologies. For mass population of people, digital contact tracing offers unparalleled advantages in monitoring their health status, quickly establishing the movement trajectories of infected people and generating the list of potentially contacts, making it one of the most important technological means in the current contact tracing for combating COVID-19. Privacy issues have undoubtedly become the major user considerations when users are deciding whether to adopt contact tracing technologies or not. They are also the current major obstacles hindering the adoption of contact tracing technologies in many countries. In response,
    %%%comment 3.12
    various protocols for preserving or protecting users privacy have been proposed and used in digital contact tracing technologies to solve or mitigate the privacy issues involved. 
    
    \subsection{Privacy Issues in Data Sharing and Usage}
    %%%增加引导语言
    The privacy concerns are raised primarily in two major aspects in relation to users data in contact tracing technologies, i.e., data sharing and data usage.
    %%%下面往上挪
    \subsubsection{Data sharing}
    
     When using contact tracing systems or apps, users are supposed to be given the privileges in controlling how their personal data are shared. At any time of the process, users should have the access to consent withdrawal which refers to a user's rights to stop sharing his personal data to any other parties. This offers users with the assurance that they can delete their data or eliminate their presence from the contact tracing app ecosystem at any time.  This privilege will help lessen users' worry about their personal data being eventually end up with untrustworthy or even malicious parties. This will also encourage users to comfortably share their data through trustable technical means with the parties involved in the contact tracing technologies.

    Specifically, the privilege of consent withdrawal should be implemented at different stages when users are using contact tracing systems or apps: 

    (1) When user data have not been transferred from the local tracing devices to the server, users should be able to delete the systems or apps from their devices, along with all the user data that have been collected and stored locally on the devices; 

    (2) If user data have already been uploaded to the server, in the case of a centralized architecture, the server should delete the user data as soon as it processes the data and, if the uploaded data contain the valuable information about the close contacts of an encounter for a positive case, alerts will also be sent out to the close contacts. In the case of a decentralized architecture, the process of consent revocation is more complex. If an infected user requests their personal data to be deleted, the server can delete the stored data it receives from the user, but it might be difficult to ensure that the data which have been transferred to other devices can be instantly removed from those devices. The possible solution is to ensure that the transferred data are always deleted after a specific time duration (e.g., 21 days) from all the devices involved in the architecture. On top of that, enhanced data encryption among all the nodes in the architecture can be implemented as the additional protective measures just in case the shared data are not deleted in time. 

    (3) Once the system or app is deactivated or phased out at the end of a pandemic, the collected user data must be automatically deleted even though no user consent revocation is received, unless the user has explicitly agreed that their stored data be migrated to other locations.

    % Most of the major tracking software on the market today are either centralized or distributed architectures. Based on these two main architectures, privacy protection can be approached from several perspectives.
    %%%目前已有混合制的啦
    % How to complement the strengths of these two architectures to build a hybrid contact-tracking ecosystem may be a pressing issue for future developers.
        
    \subsubsection{Data usage}
        
    % The question of whether user data may be used for purposes other than tracking contacts is also a perspective from which to assess the system. 
    We understand that the location information of users might be misused for the purposes other than contact tracing. For example, the information may be used by commercial entities to push advertisements to potential customers or by adversaries to launch various fraudulent or criminal activities, seriously threatening the privacy and rights of users. As a strict rule, these data can only be used in response to the COVID-19 pandemic, and other purposes of usage are strictly prohibited. Transparency in the use of data is the top priority to ensure public trust in contact tracking systems and apps. 
    Furthermore, users should be well informed of how their data are being used. The guarantee from contact tracing systems or apps in deleting user data after the tracing is completed alone is far from being sufficient. Users should be advised about what part of their data are used for what purpose in contact tracing at what time. Contact tracing systems and apps should receive explicit consent from users before their personal data can be used.
    
    \subsection{General Privacy Principles}
    Contact tracing applications should be designed with the following principles regarding privacy in mind \cite{ch06}.

\begin{itemize}
    \item \textbf{Legitimacy.} Contact tracing applications must comply with all applicable laws, rules and regulations;
    \item \textbf{Informed consent.} Contact tracing applications can only be installed and used if the user's informed consent to each feature of the application is ensured;
    \item \textbf{Identity control.} Users make the determination to release redacted, disconnected, and/or aggregated space-time points from location data;
    \item \textbf{Transparency.} Users should be informed of an organization's information practices before any personal information is collected from them;
    \item \textbf{Accountability.} Relevant stakeholders should be fully involved and consulted in the development and deployment of contact tracing applications, including data protection authorities, the privacy and security community, human rights and civil liberties organizations, government agencies, the technical community and public health professionals, including epidemiologists;
    \item \textbf{Voluntary compliance.} The use of contact tracing applications should be voluntary, not mandatory or compulsory.
\end{itemize}

The following is the checklist of some detailed considerations that takes into account the aforementioned privacy issues and principles when contact tracing systems and apps are developed or evaluated. 

\begin{itemize}
    \item Whether to limit the amount of data that can be publicly released; 
    \item Whether to provide tools to allow the diagnosed person and their healthcare provider to redact any sensitive locations, such as home or work; 
    \item Whether to encrypt location data end-to-end before sensitive locations are redacted;
    \item Whether to eliminate the risk of third-party access to information by allowing the infected person to voluntarily self-report;
    \item Whether to support the collection of data around any entities, particularly the government, to strictly regulate access to and use of the data;
    \item Whether targeted, affirmative, informed consent is obtained each time personal data is used;
    \item Whether users are provided with the ability to see how their data is being used and to withdraw consent for the use of their information;
    \item Whether users are provided with the ability to correct erroneous information.
    \item Whether individuals are informed about which data, how long it is stored, and who has access to it at each stage of use;
    \item Whether user location data is deleted promptly after it is no longer necessary to perform contact tracing;
    \item Whether open source software is used to foster trust in the application's privacy claims.
\end{itemize}

    \subsection{Architectures and Protocols}
   % As mentioned above, there are a lot of privacy issues using proximity tracing tools , and many protocols have been proposed to deal with some issues.

    According to what data is used or stored by the protocol, we can classify the architecture of protocols into three classes: centralized, decentralized and hybrid. In this section, we discuss these three architectures and relevant protocols. Table \ref{tab:Protocols} summarizes the protocols described in this survey. 
            
    \begin{table}
      \centering
      \caption{Protocol information summary}
      \label{tab:Protocols}
      \begin{tabular}{p{2cm}p{3cm}p{2cm}p{4.7cm}}
        \hline
         \textbf{Architectures}&\textbf{Name}&\textbf{Release Time}&\textbf{Country/Team}\\
        \hline
        Centralized & \textbf{Bluetrace} & Apr 10, 2020 & Singapore Government\\
		 &\textbf{ROBERT} & Apr 18, 2020 & Inria and Fraunhofer\\
		  &\textbf{PEPP-P} & Apr 1, 2020 & An European coalition of echnologists and scientists \\
        Decentralized & \textbf{PACT (East-coast)} & Apr 8, 2020 & MIT Computer Science and CSAIL \\
		&\textbf{PACT (West-coast)} & Apr 7, 2020 & A team from the University of Washingto\\
		&\textbf{Whisper Tracing} & Apr 16, 2020 & Nodle\\
		&\textbf{DP-3T} & Apr 4, 2020 & École Polytechnique Fédérale de Lausanne, ETH Zurich, KU Leuven, Delft University of Technology, University College London, Helmholtz Centre for Information Security, University of Torino and ISI Foundation\\
		&\textbf{GAEN} & Apr 10, 2020 & Apple and Google \\
		&\textbf{TCN} & Mar 17, 2020 & TCN Coalition, Covid Watch and CoEpi\\
% 		&\textbf{Bychain} & Jul 2, 2020 & Beijing University of Posts Telecommunications\\
		
        Hybrid & \textbf{DESIRE} & Aug 4, 2020 & PRIVATICS Team in Inria, France \\
		&\textbf{ConTra Corona} & Apr 29, 2020 & Germany\\
		&\textbf{EpiOne} & Apr 28, 2020 & The University of California at Berkeley\\
      \hline
    \end{tabular}
    \end{table}
    
    \subsubsection{Centralized Architectures} 
    
    In a centralized architecture, there is a central database used to store users' history data. In centralized contact tracing, location and contact data are collected and integrated centrally by a single agency. Additional information about the users, such as mobile communications service provider or payment data, may be collected and paired with location data. Central authorities identify infected people, determine their contacts, and request specific actions from those who may have been exposed to the virus. The central system can create powerful tools for analysis and public health decision-making. However, such systems also expose personal data to central authorities, creating risks in undermining individual privacy.
    
    \paragraph{Bluetrace protocol} Bluetrace protocol \cite{jt3} is developed by the Singaporean Government for logging Bluetooth encounters between participating devices to power the contact tracing for the TraceTogether app, and at the same time, protecting users' personal data and privacy. Bluetrace protocol works in the following several steps: 
    
    %%%comment 3.14    
    \begin{itemize}
    \item Users register the app using their mobile phone number, and then the back-end service generates randomly a unique ID, and bind with each user's phone number. The phone number is required for registration and can be used to notify the exposed users;
    
    \item Users exchange their information via Bluetooth to record each encounter. The information is some temporary IDs (TempIDs)  produced every 15 minutes or at other specified frequency. As shown in Figure \ref{TempIDs}, tempIDs includes userID, start time, expiry time, IV and Auth Tag. The message is encoded and won’t reveal personal information, but Healthy Authority can obtain the contact information from it.
    
    \begin{figure}[h]
      \centering
      \includegraphics[width=10cm]{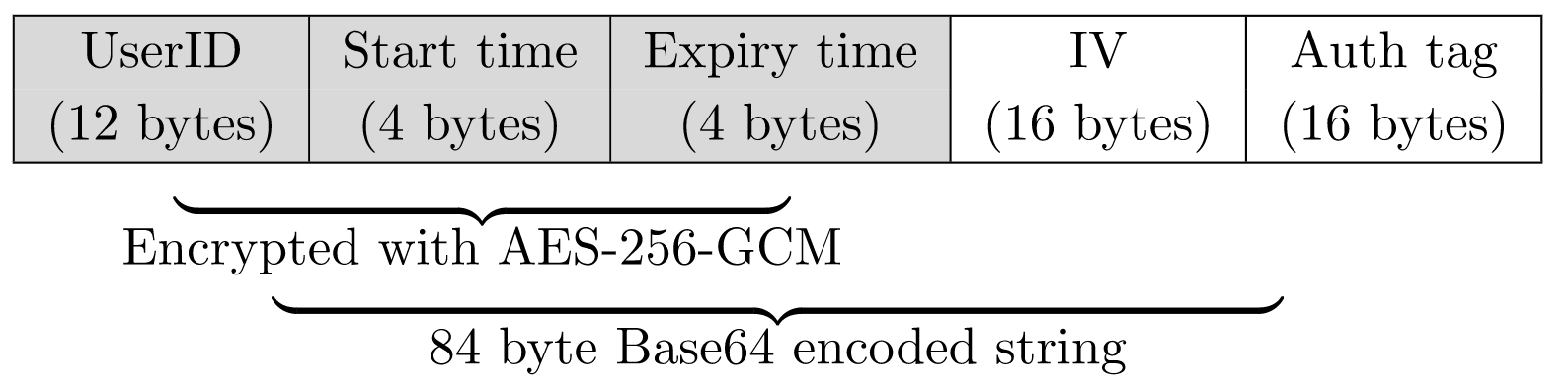}
      \caption{Format of TempIDs \cite{jt3}.}
      \label{TempIDs}
    %   \Description{temporary IDs.}
    \end{figure}
    
    \item Users' history data will be stored on the server for a certain number of days (e.g., 25 days);
    
    \item When an user has been infected, Healthy Authority will ask him/her to upload the encounter history data;
    
    \item According to the time and distance of contacting, Healthy Authority will find the close contact and notify them;
    
    \item Users have the option to deny access to their data by the server. If so, then the Healthy Authority should delete the data related to them on the back-end server.
    \end{itemize}
        
    \paragraph{ROBERT} ROBust and privacy-presERving proximity Tracing protocol \cite{jt8} is jointly developed by the researchers at Fraunhofer in Germany and INRIA in France. It has basically the same principle as the Bluetrace protocol. The main difference lies in that the data stored on the ROBERT server are anonymous identifiers called Ephemeral IDs (EphIDs). In the notification step, users need to frequently check their used EphIDs to determine if they are exposed to infected people.

    %%%%PEPP-PT也是一个protocol
    %%%comment 3.15
    \paragraph{PEPP-PT} The Pan-European Privacy-Preserving Proximity Tracing (PEPP-PT) \cite{jt9} is developed by an European coalition of technologists and scientists from over eight countries, and led by Germany’s Fraunhofer Heinrich Hertz Institute for telecommunications (HHI). It uses Bluetooth Low Energy (BLE) to discover and locally log individuals near a user, and ROBERT is a proposal for PEPP-PT protocol.

    \subsubsection{Decentralized Architectures} 
            
    In a decentralized architecture, all personal data is stored locally on users' devices, and people voluntarily decide whether to upload their data. Only location data of people identified as infected needs to be shared. 
    % An example of a decentralized approach is an Israeli application called "Tracking Viruses", as is the COVID Secure Path. 
    \par Decentralized systems typically provide greater privacy protection and are therefore more compliant with privacy requirements and regulations such as GDPR \cite{jz01}. Tools such as culling and obfuscation of infected people's data can be used to help protect their privacy. Some utility may be lost compared to a centralized system, as large datasets of users collected and aggregated can be used for useful public health research. However, when we consider various approaches, the serious privacy risks associated with centralized systems often far outweigh the limited additional benefits, leading us to place a high value on decentralized approaches. 
            
    \paragraph{PACT (East-coast)} The Private Automated Contact Tracing protocol \cite{jt10} is developed by the Computer Science and Artificial Intelligence Laboratory (CSAIL) at Massachusetts Institute of Technology (MIT). It includes two major layers:
        
    \begin{itemize}
    \item \emph{Chriping layer}. Each smartphone will generate a 256-bit random seed every hour, and every few seconds, it will generate and broadcast (using Bluetooth) a 28-bit ‘chrip’ value which combines the random seed with the current time. The ‘chrip’ value is anonymous and random, thus won’t reveal any information about users;
    
    \item \emph{Tracing layer}. All metadata (e.g., received signal strength) will store on device locally until the user has tested positive. When a user is diagnosed positive, he will be authorized to upload his logs. People who have been in contact with him should read the metadata information to check if they have really been in close contact with him;
    
    %\item \emph{Interacting with medical professions layer}. The process mentioned above is just a tool to inform individuals if they have been in the risk of exposure, anyway, it is up to doctors and health authorities to decide if individuals should be tested.
    \end{itemize}
        
\begin{figure}[h]
      \centering
      \includegraphics[width=12cm]{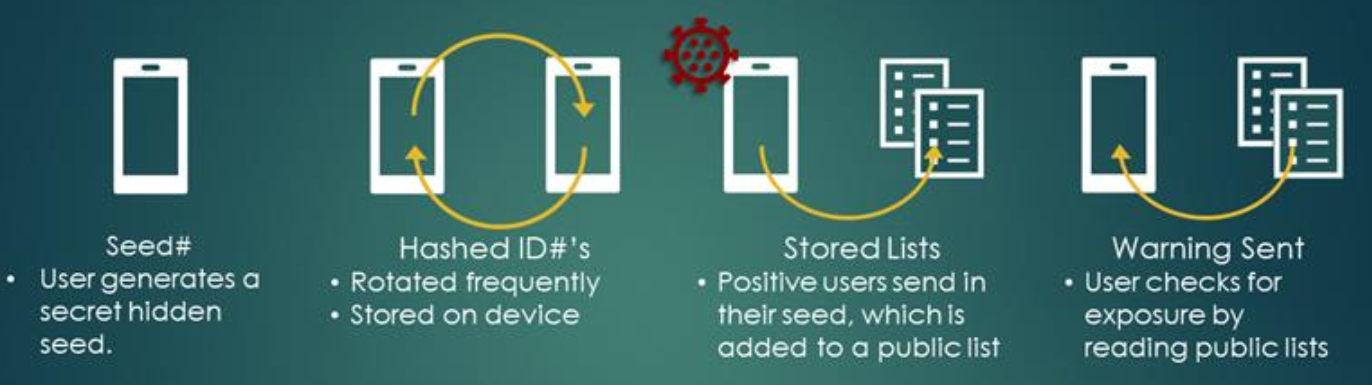}
      \caption{The process of PACT (West-coast) \cite{jt11}.}
      \label{decentralised architecture}
    %   \Description{Privacy sensitive protocols And mechanisms for mobile Contact Tracing.}
    \end{figure}

    \paragraph{PACT (West-coast)} Privacy sensitive protocols and mechanisms for mobile Contact Tracing \cite{jt11} is developed by a team from the University of Washington. As is shown in Figure \ref{decentralised architecture}, PACT (West-coast) is similar to PACT(East-coast). The main difference comes from the data it broadcasts. PACT (West-coast) uses a secure cryptographic \emph{pseudorandom generator} like $G(x)=SHA-256(x)$ to generate pseudorandom IDs. The user initially samples a random \emph{n}-bit (take 128 for example) seed, the generator generates a 256-bit length output, 128-bit of which is the temporary pseudorandom ID and the other 128-bit is the seed of next \emph{pseudorandom generator}. 
            
    \paragraph{Whisper Tracing.} Whisper Tracing \cite{jt24} is a decentralized and proximity-base contact tracing protocol proposed by Nodle. When running, the library locally generates temporary IDs and uses Bluetooth Low Energy (BLE) to advertise those IDs and detect proximity event with other whisper users. It is being used for the Government of Senegal's Daancovid19 mobile contact tracing app (Coalition) initiative.
    %Coalition: https://www.coalitionnetwork.org/
           
    \paragraph{DP-3T} Decentralized Privacy-Preserving Proximity Tracing (DP-3T) \cite{jt13} is developed by researchers from EPFL, ETHZ, KU Leuven and so on in response to the COVID-19 pandemic to facilitate digital contact tracing of infected participants. It is very similar in functionality to PACT (East-coast) discussed earlier. Users' smartphone with apps continually broadcasts an ephemeral and pseudo-random ID, in the meanwhile, records the IDs observed from other users in close proximity. The data of an user is stored on his apps until he is diagnosed with COVID-19, in which situation he can upload all of her pseudo-random IDs to a central server. So far, apps using this protocol have covered many countries, such as Switzerland, Germany, Estonia, and so on.
    % Austria, Belgium, Croatia
    %%%comment 3.13
    \paragraph{GAEN} Google/Apple exposure notification protocol is developed by Apple and Google. Similar to other protocols including PACT and DP-3T, the GAEN protocol uses Bluetooth to detect proximity with others. Figure \ref{GAEN} shows the overview of the GAEN protocol. Later, David Culler et al. enabled the GAEN protocol to support manual contact tracing efforts, provide visibility into the spread of disease, and give the authority back to the local communities while preserving privacy within the Apple and Google framework:
   
    % the lighthouse and the covid-commons
            
    \begin{figure}[h]
      \centering
      \includegraphics[width=\linewidth]{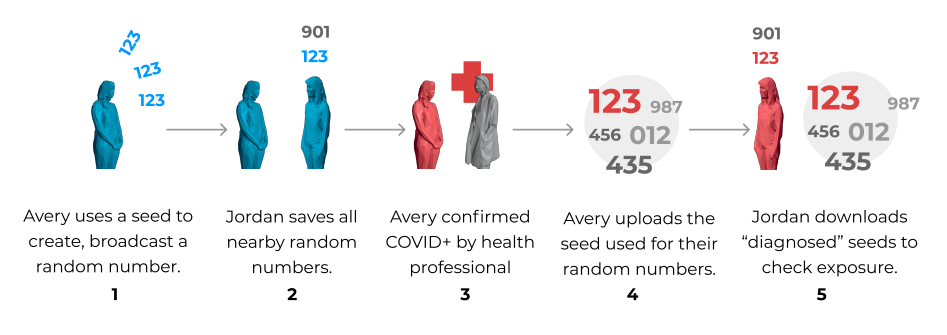}
      \caption{The process of GAEN \cite{jt14}.}
      \label{GAEN}
    %   \Description{Google/Apple exposure notification protocol overview.}
    \end{figure}

    \paragraph{TCN}Temporary Contact Number protocol \cite{jt15} is developed by TCN Coalition, Covid Watch and CoEpi in response to the COVID-19 pandemic. Applications following this protocol use iOS and Android apps' capability to share a 128-bit Temporary Contact Number (TCN) with nearby apps using Bluetooth Low Energy (BLE).
    % 后面有空再挖!!!
            
    % \paragraph{Bychain} Bychain \cite{jt16} is a decentralized and permissionless blockchain protocol for Privacy-Preserving Large-Scale Contact Tracing, which was developed by researchers in Beijing University of Posts Telecommunications, China. The protocol protects data security and location privacy, while efficiently and dynamically deploying SRC Internet of Thing (IoT) witnesses to monitor large areas.
    
    % 后面有空再挖!!!
            
    \subsubsection{Hybrid Architectures}
    In the centralized architecture, almost all tasks are completed in the back-end server, e.g., TempID generate and receive, risk analysis and notifications for the close and at-risk contacts. On the contrary, all the tasks are finished on devices like mobile phone with apps in the decentralized architecture, and the back-end server is just used as a bulletin board for finding close contacts. The hybrid architecture has combined the features of the centralized and decentralized architectures to achieve the advantages of both architectures. The tasks are split between the back-end server and the devices. Specifically, TempID generation and management takes place on the devices to protect users' privacy, while risk analysis and notification takes place on the back-end server. 

    \paragraph{DESIRE}DESIRE \cite{jt17} is developed by PRIVATICS Team in Inria. It has combined the advantages of the protocols based only on the centralized or decentralized architecture. It works based on the following ideas:
            
            \begin{itemize}
            \item Users register on an APP and the back-end server will send an unique ID to the APP;
            \item Mobile phones will generate their own Ephemeral ID (EphID) every 15 minutes. When user encounter, the receiver will generate and store some PETs to record the encounter. The process of PETs generation is shown in Figure \ref{PETs}.
            \begin{figure}[h]
              \centering
              \includegraphics[width=8cm]{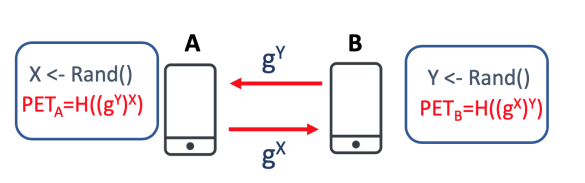}
              \caption{PETs Generation \cite{jt17}.}
              \label{PETs}
            %   \Description{The process of generate PETs \cite{jt17}.}
            \end{figure}
            
            \item When a user has tested positive for COVID-19, he will be required to upload the data, including ID, PETs and so on. Anyone who wants to check if he has the risk of exposing to an infected individual, he should upload his PETs data and the server will make a risk analysis by comparing the PETs from the infected individual.
            \item If a user finds that his score of the exposure risk is high, then he should contact the Healthy Authority and ask for advice.
            \end{itemize}
            
    \paragraph{ConTra Corona} ConTra Corona
    %(Contact Tracing against the Coronavirus by Bridging the Centralized–Decentralized Divide for Stronger Privacy)
    \cite{jt18} is developed by German researchers from the FZI Research Center for Information Technology and Karlsruhe Institute of Technology. 
    %%%comment 3.18
    Different from DESIRE, it employees three different servers: the Submission Server, the Matching Server and the Notification Server.
    \begin{itemize}
    \item Devices will generate a warning identifier (\emph{wid}) everyday based on the user’s real identity and store it for later use, deleting it after four weeks. For each \emph{wid}, devices will compute 96 \emph{sids} and the ephemeral public identifiers \emph{pid} by using the encryption algorithm. All users' (\emph{sid},\emph{pid}) pairs will be uploaded to the Submission Server.
    \item The Submission Server will send these (\emph{sid},\emph{pid}) pairs to the Matching Server after shuffling them. Once the Matching Server recieve a \emph{pid} from the inflected user, it will look up all \emph{sids} that potentially contaminated users, and send them to the Notification Server.
    \item The Notification Server will recover users' \emph{wid} according to their \emph{sid}, and publish the \emph{wid} list. Users can get the \emph{wid} list from the Notification Server and compare with the \emph{wids} they have used in four weeks.
    \end{itemize}

\begin{figure}[h]
              \centering
              \includegraphics[width=13cm]{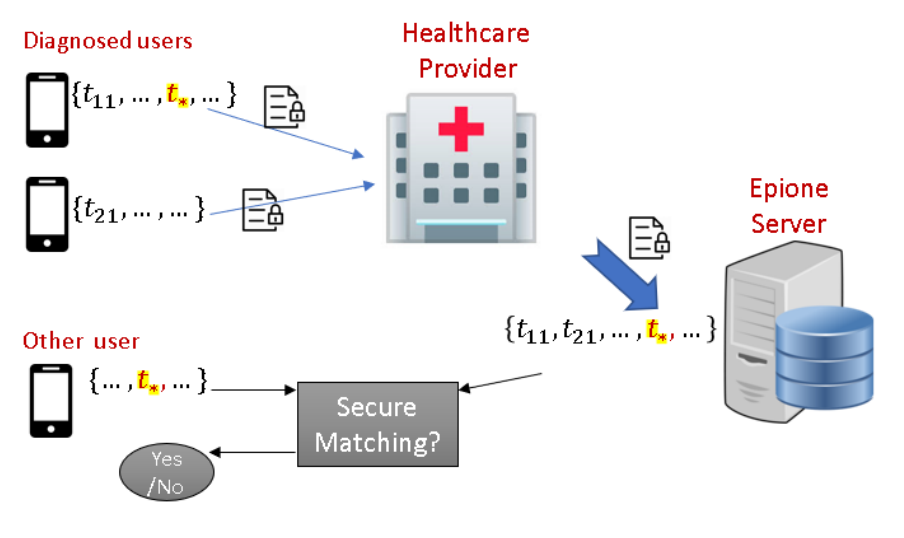}
              \caption{An overview of EpiOne \cite{jt19}.}
              \label{Epione}
            %   \Description{EpiOne overview.}
            \end{figure}

    \paragraph{EpiOne} Epione (Lightweight Contact Tracing with Strong Privacy) \cite{jt19} is developed by a research group led by the University of California at Berkeley. Similiar to TCN protocol, it uses a short-range network (such as Bluetooth) to detect when two users are within a close range and exchange a randomly generated "contact token". Figure \ref{Epione} shows an overview of EpiOne.
            
\section{Digital contact tracing methods}
    Digital contact tracing has existed as a concept since at least 2007 \cite{digital_his1} \cite{digital_his2} and it was proven to be effective in the first empirical investigation using Bluetooth data in 2014. The concept came to prominence during the COVID-19 pandemic, where it was deployed on a wide scale for the first time through multiple governments and private COVID-19 apps. 
    Considering its generic principles and algorithms involved, digital contact tracing can also be applied to other respiratory infectious diseases as well.
    
    \subsection{The General Architecture of Digital Contact Tracing}
\begin{figure}[h]
      \centering \includegraphics[width=\linewidth]{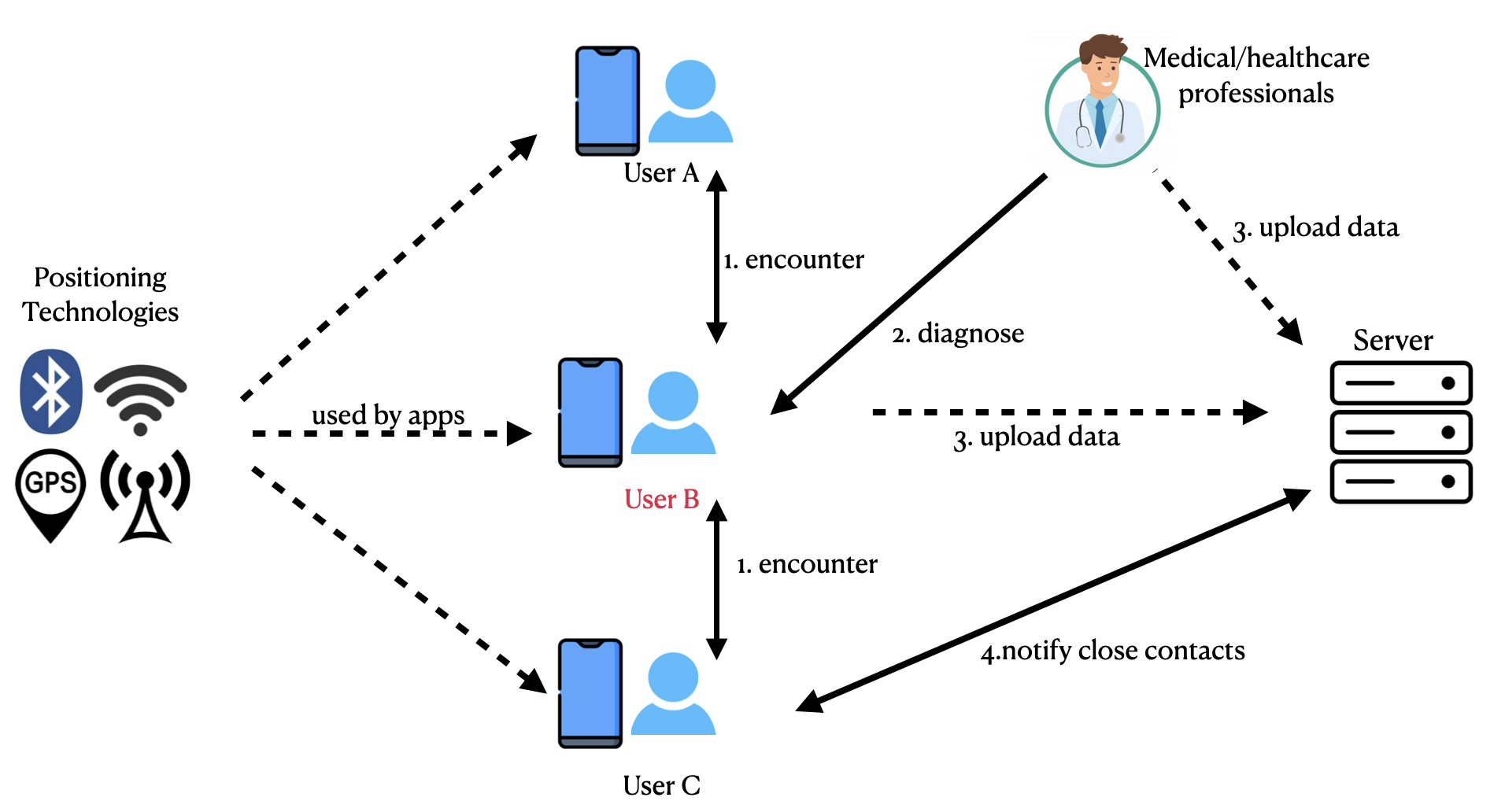}
      \caption{The General Architecture of Digital Contact Tracing Apps.}
      \label{digital contact tracing process}
    %   \Description{}
    \end{figure}

    Digital mobile devices, such as smartphones, have been deemed as the most ideal tracking devices for contact tracing thanks to their popularity in use by the general public. Using smartphones as the platform, a scores of mobile apps have been developed by many different countries for dedicated contact tracing use. Even though those apps may be quite different in terms of the localization technologies, data storage and communication protocols used, their general working mechanism is highly similar. Figure \ref{digital contact tracing process} presents the general architecture for the dedicated mobile contact tracing apps developed on smartphones. 
    
   The localization technologies that can support various contact tracing apps include Bluetooth, GPS, Cellphone tower networks and WiFi. Contact tracing apps typically use one or two of those localization technologies to acquire proximity or encounter information, which is the important information required in contact tracing. 

    There are two groups of users in these contact tracing apps, i.e., the general public (represented by User A, B and C in Figure \ref{digital contact tracing process}) and the medical/healthcare professionals who are responsible for diagnosing patients and carrying out contact tracing. Here, we suppose that all users involved (User A, B and C) have installed and registered for using the contact tracing app, and User B is tested positive to the infectious disease, e.g., COVID-19.

The general process of digital contact tracing using mobile apps takes the following four steps:
    
    \begin{itemize}
        \item Every time when users make encounters with each other, the related information (e.g., the location and encounter information) will be exchanged and stored in the apps of users or sent to the back-end server;
        \item After User B has been diagnosed as having contracted the disease, the diagnosis information will be uploaded to the server by User B himself or by the medical or healthcare professionals;
        \item After the server receives the information of a new patient (User B), it will analyze the patient's location and encounter information to find out potential contact (i.e., User C) based on some predetermined rules in terms of the proximity and duration of contacts. Then, the app will send a message to notify User C. Necessary medical tests and/or quarantine measures will be taken for User C. Even though User A has encountered with User B as well, the encounter does not satisfy the predetermined rules, so User A is deemed to have a low risk and thus won't be notified.
    \end{itemize}

    \subsection{Contact Tracing Mobile Apps}
   A plethora of mobile apps have been developed for digital contact tracing during the COVID-19 pandemic to identify more efficiently the persons who may have been in contact with an infected individual, but their effectiveness is heavily dependent on their install rate. In this subsection, we will detail a number of popular contact tracing mobile Apps, that have been adopted by different countries in contact tracing of COVID-19. 
   Similar to the protocol, these contact tracing apps can generally be classified into two classes by whether the contact tracing is done on a centralized server or done on every user's phone in a decentralized way.
   %%%comment 3.7
   Table \ref{Mobile Applications} presents a summary of these apps, which all can be downloaded from Google Play, the Apple App Store or online.

% Section 3.2 is very informative but lack of a systemical approach to categorise all the apps presented (plus pros and cons).  Privacy and centralised/decentralised further discussed in sections 4.3 and 4.4 should be integrated as categorisation principles in that section 3.2 (as well as most of the further issues discussed in the remaining sections). 
% Next all the applications presented are not discussed in terms of their success/failures, a feedback on the way they are successful or not will be of interest here.

%%%comment 3.8
\begin{table}
      \centering
      \caption{Summary of Mobile Apps for COVID-19 Contact Tracing.}
      \label{Mobile Applications}
      \begin{tabular}{p{2cm}p{4cm}p{1.6cm}p{1.6cm}p{3cm}}
        \hline
         \textbf{Architectures} & \textbf{Name} & \textbf{Release Time} & \textbf{Country} & \textbf{Protocol}\\
        \hline
         Centralized & \textbf{Trace Together} & Mar, 2020 & Singapore & BlueTrace\\
         & \textbf{CovidSafe} & Apr, 2020 & Australia & BlueTrace\\
         & \textbf{Aarogya Setu} & Apr, 2020 & India & GPS \ Bluetooth\\
         Decentralized & \textbf{COVID Tracker Ireland} & Jul, 2020 & Ireland & ENS and Bluetooth\\
         & \textbf{HaMagen} & Mar, 2020 & Israel & GPS\\
         & \textbf{Covid Watch} & Apr, 2020 & Arizona & GAEN API \\
         & \textbf{SwissCovid} & Jun, 2020 & Switzerland & DP-3T\\
         & \textbf{Corona-Warn-app} & Jun, 2020 & Germany & GAEN API\\
         & \textbf{HOIA} & Aug, 2020 & Estonia & DP-3T \& GAEN API\\
         & \textbf{Coalition} & Apr, 2020 & Senegal & Whisper Tracing \\
        % \textbf{QR Code based apps} & Feb, 2020 & China & GPS\\
        % \textbf{Aura} & July, 2020 & US & QR\\
        % \textbf{SCT system} & May, 2020 & US & Bluetooth\\ 
      \hline
    \end{tabular}
    \end{table}
   %%%comment 1.2 secondly
   \subsubsection{Centralized Architectures}
   
   As mentioned in protocols, in centralized architectures, all contact tracing tasks are done in a centralized servers. After infected users upload their data, central Health Authorities can notify their contacts. In this kind of architectures, app developers, health workers and Health Authorities may know who have installed the app, who have been infected and who may have been exposed to the virus. In this way, Health Authorities can inform the exposed users and request specific actions from them. However, users will only know whether they have been exposed to the virus recently.

        \paragraph{TraceTogether}
        
        The TraceTogether app \cite{TraceTogether} is a Bluetooth-based app proposed by the Singapore Government in March 2020. It uses a digital contact tracing protocol called BlueTrace. Phones exchange anonymized proximity information using Bluetooth during encounters. This information is stored securely on the phone and is automatically deleted after 25 days. If a user tests positive to COVID-19, the data is only shared with the Ministry of Health (MOH) . 
        Using received signal strength indicator (RSSI) values, TraceTogether can measure the signal strength between devices. Calibrated RSSI values are used to estimate the approximate distance between users during an encounter. TraceTogether interpolates between successive communications to estimate the approximate duration of an encounter.
        
        Singapore's National Development Minister, Lawrence Wong, has said, if we want the tracing technology to be truly effective, three-quarters of country's residents must have the app installed on their mobile phone. But a month after TraceTogether's launch, there were only 1.08 million users (5.7 million residents total) in Singapore, which is far from the target \cite{jt30}. Due to the app's data privacy policies, no further results are released. However, the app has attracted global interest- the governments of more than 50 nations worldwide have expressed interest in adopting or modifying the app to suit their countries \cite{jt36}.
      
        % \begin{figure}[h]
        %   \centering \includegraphics[width=7cm]{./TraceTogether.png}
        %   \caption{TraceTogether (ios version) \cite{TraceTogether_app}.}
        %   \label{TraceTogether}
        %   \Description{}

        % \end{figure}

        \paragraph {CovidSafe}
        
        CovidSafe \cite{COVIDSafe} was released by the Australian Government on April 26, 2020, with the source code for both Android and iOS clients released on 8 May 2020 \cite{COVIDSafe_code}. On August 1, 2020, NSW Health announced the app had helped them trace new contacts. They accessed the app data on a corona-virus case and identified 544 additional people, two of whom are tested positive to COVID-19 \cite{COVIDsafe_report}. Covid-Safe is based on the Bluetrace protocol and exhibits many similar characteristics to the TraceTogether app. The app augments traditional contact tracing by automatically tracking encounters between users, and later allowing Healthy Authority to warn a user if he has come within 1.5 meters with an infected patient for more than 15 minutes.
        
        A week after CovidSafe's launch, there were more than 4 million users (25 million residents in Australia total) had downloaded the app. In the first two weeks, it had more than 5 million downloads and then the download rate began to reach the plateau. At June 6, 2020 the total number of people in Australia who had downloaded the CovidSafe app had reached 6.13 million \cite{jt37}.
        
        % \begin{figure}[h]
        %   \centering \includegraphics[width=9cm]{./CovidSafe.png}
        %   \caption{CovidSafe (ios version) \cite{COVIDsafe_app}.}
        %   \label{CovidSafe}
        %   \Description{}
          
        % \end{figure}
        
        \paragraph {Aarogya Setu} 
        
        Aarogya Setu \cite{Aarogya_site} is an Indian open-source COVID–19 contact tracing, syndromic mapping and self-assessment digital service. It was developed by the National Informatics Centre under the Ministry of Electronics and Information Technology (MeitY). The source code is available for Android and iOS app \cite{Aarogya_code}. It uses GPS and Bluetooth information to track corona-virus infections. With Bluetooth, it determines the risk if one has been within six feet of a COVID–19-infected person by scanning through the database. Using location information, it determines whether the location where one is at belongs to one of the infected areas based on the data available. Aarogya Setu has several functions: 
        \begin{itemize}
            \item Analyze the risk of contracting COVID-19 for the user;
            \item Helps users self-assess COVID-19 symptoms;
        \item Updates local and national COVID-19 cases;
        \item Advises the number of COVID-19 patients in a radius of 500 m, 1 km, 2 km, 5 km and 10 km from the user, respectively.
        \end{itemize}
        
        The app has played a key role in Indian contact tracing, helping identify more than 10 million people who are at risk. Using the app, high-risk individuals can be identified at an early stage so that they can be prioritized for testing and necessary medical and administrative measures. Until now, Aarogya Setu has been downloaded more than 170 million times. Chandigarh top the list with 48\% downloads and the national capital Delhi is second with 46\% downloads. Similarly, the larger States like UP, Madhya Pradesh and Rajasthan have registered only 10\%, 8.7\% and 9.2\% of the downloads of their respective populations \cite{jt38}.

        % \begin{figure}[h]
        %   \centering \includegraphics[width=9cm]{./Aarogya.png}
        %   \caption{Aarogya Setu (iOS version) \cite{Aarogya_app}.}
        %   \label{Aarogya Setu}
        %   \Description{}
        % \end{figure}
        %%%comment 1.2 secondly
        \subsubsection{Decentralized Architectures}
        On the contrary, all data and tasks are stored and executed locally on users' devices in decentralized architectures. The contact tracing app can inform the exposed user and guide them on how to test and self-quarantine.
        Apps based on decentralized architectures provide greater privacy protection, but a survey \cite{jt35} found that, most people on Amazon Mechanical Turk (based on a sample from the US only) preferred to install apps with centralized architectures for they don't want to give tech-savvy users the opportunity to infer the identity of the infected user they have been in contact with.
        
        \paragraph {COVID Tracker Ireland} 
        
        COVID Tracker Ireland \cite{Ireland_site} is a digital contact tracing app released by the Irish Government and the
        %%%comment 3.9
        Health Service Executive (HSE) on July 7, 2020.
        %%%comment 3.10
        The app uses Exposure Notifications System (ENS), developed by Apple and Google, and Bluetooth technology to determine whether a user have been in close contact with someone for more than 15 minutes who tested positive to COVID-19.
        It has four main functions:
        \begin{itemize}
            \item Alert users who are in close contact with someone who has tested positive;
            \item Alert close contacts of users who tested positive;
            \item Give users advice when they report symptoms;
            \item Provide an overview of national and regional data, e.g., confirmed cases, hospital and ICU admissions.
        \end{itemize}
        
        As of August 10,2020, over 1,560,000 people have downloaded the app and alerts for close contacts issued to 137 people \cite{Ireland_news}. Until now, about 2.4 million people in Ireland (4.9 million residents total) have signed up for the app, with an active user base of 1.3 million, the usage is relatively high. However, Ireland confirmed a total of 189,851 cases of COVID-19 on January 26, 2021, only 13,136 users with positive tests shared their results on the app, and only 21,757 users received a close contact alert \cite{jt39}. So even if the up-take is higher than the rest of the world, the app's impact is likely to be limited.
        
        % \begin{figure}[h]
        %   \centering \includegraphics[width=9cm]{./Ireland_app.png}
        %   \caption{COVID Tracker Ireland (iOS version) \cite{Ireland_app}.}
        %   \label{COVID Tracker Ireland}
        %   \Description{}
        % \end{figure}
        
        \paragraph {HaMagen} 
        
        HaMagen is developed by Israel's Ministry of Health. The application cross-checks the GPS history of a mobile phone with the historical geographical data of identified cases from the Ministry of Health. The GPS data only stay in mobile phone and are not sent to any third parties. The data from the Ministry of Health are sent to user's mobile phone for cross-referencing. So the cross-referencing of users' GPS history with the locations of patients only happens on the mobile phone. After installation, the app compares the location data of verified COVID-19 patients with users' location data via GPS and can identify proximity between phones where it is installed using Bluetooth technology. An overlap (based on location or proximity between devices) between users and a verified COVID-19 patient can be detected. If this happens, the app will alert users by SMS.
        
        HaMagen enjoyed a promising start \cite{jt40} that, in its first few weeks, more than 1 million users (9.2million total) downloaded the app, and in the first week, 50,000 users reported that they have self-quarantined \cite{jt41}. But since late March 2020, the number of installs who actually use the app has dropped, even though more than 350,000 people have downloaded the HaMagen 2.0, only 22,000 users have kept it on their phones.
        
        % \begin{figure}[h]
        %   \centering \includegraphics[width=9cm]{./HaMagen.png}
        %   \caption{HaMagen (iOS version) \cite{HaMagen_app}.}
        %   \label{HaMagen}
        %   \Description{}
        % \end{figure}
        
        % \paragraph {\textbf{f}} \textbf {PRONTO-C2}
        
        % PRONTO-C2 contact tracing system is proposed by Researchers from the University of Salerno, Italy. It lets the devices communicate anonymously with each other while hiding these communications from the central server, averting mass surveillance.
        % this is a protocol!
        
        % \paragraph g) CoEpi is an open-source project based on TCN. It develops a decentralized, privacy-first app for anonymous Bluetooth-based exposure notification based on symptom sharing. Communities of close contacts can begin protecting themselves with CoEpi without requiring widespread adoption among the general population; there is no scale required to achieve benefit to small user groups. 
        
        \paragraph {Covid Watch} 
        
        Covid Watch \cite{CovidWatach_site} is an open-source nonprofit organization founded in February 2020. It started as an independent research collaboration between Stanford University and the University of Waterloo, and was the first team in the world that published a white paper \cite{CovidWatach_white} and developed an open-source anonymous Bluetooth exposure alert protocol - the CEN Protocol \cite{CovidWatach_code}, later renamed the TCN Protocol - in collaboration with CoEpi \cite{CoEpi_site} in early March 2020. 
        
        Covid Watch also built a fully open source mobile app for sending anonymous exposure alerts first using the TCN Protocol in April 2020 and later using the nearly identical Google/Apple exposure notification (GAEN) framework. The app uses Bluetooth to sense when one is near someone else who has installed Covid Watch. A log of instances when one user was in a close proximity to other users is stored solely on the phone. When tested positive to COVID-19, user can inform the app about the diagnosis. Covid Watch will automatically warn everyone who was in a close proximity about potential infection and encourage them to take action like getting tested or take self-quarantine.
        %Also in May 2020, Covid Watch launched the first calibration and beta testing pilot of the GAEN APIs in the United States at the University of Arizona. In Aug 2020, the app launched publicly for a phased roll-out in the state of Arizona.
        
        % \begin{figure}[h]
        %   \centering \includegraphics[width=9cm]{./Covid watch.png}
        %   \caption{Covid Watch (iOS version) \cite{CovidWatch_app}.}
        %   \label{Covid Watch}
        %   \Description{}
        % \end{figure}
        \paragraph {SwissCovid} 
        
        SwissCovid \cite{SwissCov_site} is used in Switzerland. The app use Bluetooth and Decentralized Privacy-Preserving Proximity Tracing (DP-3T). It was developed in collaboration with the École polytechnique fédérale de Lausanne and the Swiss Federal Institute of Technology in Zurich as well as other experts \cite{SwissCovid_dev}. When within Bluetooth range, the mobile phone exchanges random IDs with other mobile phones that have a compatible app installed. The random IDs are stored on the mobile phone for 14 days before being deleted automatically.
        
        If users test positive to the coronavirus, they receive a Covidcode from the cantonal authorities. The code allows them to activate the notification function in the app, thereby warning app users that came into close contact with the infected person. The identity of the person who triggered the notification is not revealed. App users receives notifications if they have spent more than 15 minutes within 1.5 metres from infected person within a 24-hour period.
        
        SwissCovid is effective in contact tracing and can help reduce COVID-19 infections \cite{jt42}. A team has calculated that non-household contacts notified of exposure by SwissCovid entered quarantine a day earlier than those notified through manual contact tracing \cite{jt44}. An evaluation of SwissCovid , published as a preprint in this February \cite{jt43}, found that the app boosted the number of people in quarantine in the Canton of Zurich by 5\%, and 17\% of these people tested positive last September.
         
        % \begin{figure}[h]
        %   \centering \includegraphics[width=9cm]{./SwissCovid.png}
        %   \caption{SwissCovid (iOS version) \cite{SwissCo_app}.}
        %   \label{SwissCovid}
        %   \Description{}
        % \end{figure}
        
        %%%comment 1.2 firstly+3.16
        \paragraph {Corona-Warn-app}
        
        Corona-Warn-app \cite{jt32} supported by the federal government, German companies SAP and Telekom and used in Germany from June 16, 2020, is based on the Exposure Notification API from Apple and Google (GAEN API) which inspired by DP-3T. It uses Bluetooth technology to exchange anonymous encrypted data from a distance of a few metres with other mobile phones and may then know from the device whether an infected person is nearby. For example, if the current owner of phone A is infected, the device can send an alert to all phones that have been around for A while and have the app installed. Other phone owners will receive a push message and can be tested \cite{jt31}.
        
        Just one week after Corona-Warn-app was launched, 15\% of people living in Germany (83 million total) have downloaded it, and experts in Oxford said that a tracing app can play a key part in any pandemic concept if it adopted by at least 15\% of the population \cite{jt45}. In July 2020, about 16 million Germans had already installed the app.
        
        \paragraph {HOIA}
        
        HOIA is devolped by a motley ensemble of Estonian IT companies and state institutions and rolled out on August 2020. It use DP-3T as the basis for the new coronavirus app, while using the GAEN API. This app relies on Bluetooth low energy technology or BLE and nearby phones using the app will exchange their unique anonymous codes with each other. When someone has been confirmed to be infected, the anonymous codes on the device will be uploaded to a central server. And the device will regularly download anonymous codes from the infected people and check if people have been in close contact with an infected person and notify people when a close contact is detected \cite{jt33}.
        
        According to the information announced by the Health and Welfare Information Systems Center (TEHIK), HOIA has been downloaded 250,944 times until to this January , and 2,416 people have marked themselves as being ill using the app, of which active cases currently number 540 \cite{jt46}.
        
        \paragraph {Coalition} 
        
        Coalition was launched in early April 2020 by a non-profit called Coalition Network. It is a privacy-first exposure notification app that use Bluetooth and Whisper Tracing.  The mobile phone will generate temporary random IDs and change with other phone passed by. When a Coalition app user declares that they are showing symptoms or have tested positive for COVID-19, their phone then will alert all other phones that have spent time or come into contact with them over an extended period of time during the prior 2 weeks. Users who are notified can then immediately take necessary self-isolation measures to stop the spread of COVID-19 \cite{jt34}.

    \subsection{Localization} Technologies
    
    In support of the various mobile apps developed in response to the COVID-19 pandemic, many different localization technologies have been used. 
    %%%comment 1.4
    In this section, we discuss several major and relatively mainstream localization technologies, including Bluetooth, GPS, Cellphone tower networks, Wi-Fi and Acoustic-ranging, that have been widely used for contact tracing. GPS can only work outdoors while Bluetooth, Wi-Fi, Acoustic-ranging can all work indoors. Table \ref{tab:localization technologies} gives a comparison among different localization technologies.

\begin{table}
      \centering
      \caption{Comparison between localization} technologies.
      \label{tab:localization technologies}
      \begin{tabular}{p{2.5cm}p{3.4cm}p{4.7cm}p{1.8cm}}
        \hline
        \textbf{Technologies} &\textbf{Data}&\textbf{Accuracy}&\textbf{Battery use}\\
        \hline
         \textbf{Bluetooth} & Record the encounter between each devices & Able to classify close contacts within 2 metres & Low loss \\
         \textbf{GPS} &Record the GPS location & Unable to filter for proximity and has a higher false positive rate & Medium loss\\
        \textbf{Cellphone tower networks} &Capture and store the information of cellphone communication & Low accuracy of the location obtained & Negligible\\
        \textbf{Wi-Fi} & Scan and store information about all surrounding devices & Very accurate if people are in the same Wi-Fi netwoeks, but it is typically limited in its range and coverage & Negligible\\
        \textbf{Acoustic-ranging} & sound communication between devices &  High accuracy but will decrease as the number of devices communicating with each other increases & Negligible\\
      \hline
    \end{tabular}
    \end{table}
    % many digital tools have been developed to assist with contact tracing and case identification. These tools include outbreak response, proximity tracing (location-based (GPS, Cell tower, or Wi-Fi, Bluetooth technology, etc.), and symptom tracking tools, which can be combined into one instrument or used as stand-alone tools \cite{jt2}. In this part, we mainly introduce the proximity tracing tool.
    
    \paragraph{a}Bluetooth
    
    Bluetooth is used between two phones and has the advantage of a lower battery usage. Using Bluetooth, we can obtain traces of contact devices with a resolution of 1-2 meters, which makes it ideal for identifying close personal contacts that are most likely to transmit infectious diseases. A major limitation of Bluetooth is that it is only able to record the proximity information of users within a small neighborhood. So Bluetooth is ideal in generating the list of contacts in contact tracing, but is not capable of producing the spatial location of users, making it difficult to establish more accurate and complete movement trajectories of people to support more advanced analysis if necessary.
    %%%comment 1.4
    In addition, an inherent limitation of Bluetooth is its poor indoor positioning accuracy, such as in factories, office buildings, hospitals, retail stores and hotels, due to absorption, line of sight blockages and multi-path reflections in indoor spaces \cite{jt29}.
    %However, it is unable to track patients who may have become infected by touching a surface an ill patient has also touched \cite{jt3}. 
   
    \paragraph{b}GPS
    
    %%%comment 1.4
    Location-based technologies based on the Global Positioning System (GPS) can track the users' location and be used to identify people who have been in the same location as the infected person, to facilitate contact identification and tracing \cite{jt2}. GPS can work generally well in the outdoor environment - the accuracy of GPS-based tracking is about 10 to 15 meters outdoors with errors 20-30 feet, but it is severely reduced indoors due to the weak GPS signals, which is a major limitation for it to contact tracing since most infectious contact occurs indoors. The usage of GPS is more private than Bluetooth, but can also lead to privacy problems \cite{jt5}. 
   
    \paragraph{c}Cellphone tower networks
    
    Cellphone tower networks can achieve location tracking. Using Cellphone tower networks don’t require users to download and use a particular app. It was first deployed in Israel \cite{jt12}. It is non-intrusive and can be put in place without any user intervention assuming a legal framework is in place. One of the major advantages of this technology is that the communication providers have been already capturing and storing the information of cellphone communication, so the information is readily available for contact tracing. The economic cost of using this technology is therefore negligible compared to other technologies. However, since the area of any cell may vary from a few hundred meters to several kilometers, the accuracy of the location obtained will be very low, and the process of determining the contact based on these traces is very inaccurate. Also, this technology could pose serious privacy concerns \cite{jt4}. 
    
    \paragraph{d}Wi-Fi
    
    Using Wi-Fi we can determine the identity (MAC address) of the surrounding devices, which help generate a very accurate list of possible contacts as long as they are in the same Wi-Fi networks. Also, mobile nodes can periodically scan and store information about all these surrounding devices, including the strength of the received signal (RSSI), which can be used to estimate distance. However, Wi-Fi may not be ubiquitously available and it is typically limited in its range and coverage. 
    %%%comment 1.4
    However, Wi-Fi is a good approach for indoor contact tracing. In enterprises, there many access points (APs), user devices constantly get connected, disconnected, and move between APs. All these activities are saved by each AP in "syslog" file. By analyzing the events logged in the "syslog" file for each device we can identify the spatio-temporal mobility trajectory of each device. Then, a contact graph will be generated for contact tracing.
    %%%comment 1.4
    \paragraph{e}Acoustic-ranging
    
    Acoustic-ranging is sound-based contact tracing with the idea that every device emits a random but unique acoustic signal that can be used to infer distances \cite{jt27}. The technology infers distance by accurately measuring the time it takes for a sound to travel from one device to another and multiplying it by the speed of sound. The subtle difference in time measurement won't cause a large error in distance, so acoustic-ranging based systems feature a high tracing accuracy. However, sound-based contact tracing can work well for the communication between two devices, but it could be very noisy when the number of devices increases. 
    %In addition, the frequency and amplitude of the acoustic signal are chosen to beyond the range of human hearing, but some animals can hear them, so it need more rigorous tests.}

\section{Other Issues of Contact Tracing Technologies}

Besides the privacy issues that we have discussed in Section 3, we identify the following several additional important issues for digital contact tracing that should also be considered when developing or evaluating contact tracing technologies. 

\subsection{Security}
The user data used in contact tracing are useful, important and often highly sensitive. The security of the contact tracing solution contributes to users' perception that their privacy are properly protected when they are sharing their personal data for the purpose of contact tracing, which will further helps boost user adoption of the solution. 

Contact tracing systems or apps must keep data secure by safeguarding the data against various internal and external security threats. A wide range of established security measures, algorithms and techniques can be implemented on the servers, tracing devices and data communication channels \cite{jt2}. For example, computer-based security measures can be used to effectively block threats, high-level firewall is able to block malicious access from the networks, and sophisticated encryption can be applied to data transmission. Extra caution should also be taken when sharing information to any third parties even if this has been agreed by users. Investigations on the level of the third parties in securing sensitive data might be necessary. 

\subsection{Transparency}

Adopting the open-source principal to make the source code of conduct tracing systems or apps publicly available is an effective way to achieve transparency and trust, which are of great importance to wider adoption by end users. Thanks to the transparency in the underlying working mechanism given to external parties, user trust can be better established towards the developers, and the service providers in data communication and storage involved in the systems or apps. Also, it would be useful to ensure regular reviews of code and credible third-party audits. This allows external parties outside the development team to review the project's code to fully understand the internal working mechanism of the technology, including the privacy and security features implemented. The external parties can also make improvements to fix the existing loopholes or bugs, or use the source code as the basis to develop new features at/or new systems or apps. 

Nevertheless, we also understand that, in some occasions, the application development team can rightfully choose not to publish their codes due to some valid reasons, e.g, they are seeking to develop some proprietary technologies or using some patented technologies that are under IP protection.

\subsection{Tracing Effectiveness}
When developing and evaluating contact tracing technologies, we hope that they are effective in quickly identifying the chain of virus transmission in order to contain the infection. Yet, we should understand that the effectiveness of contact tracing technologies can be affected by many factors. Next, we elaborate on the factors from three major perspectives:

\subsubsection{Implementation of scientifically correct contact tracing rules.} Contact tracing rules are developed based on domain expert knowledge and need to be implemented in contact tracing technologies. Given what is currently known about COVID-19 human-to-human transmission, contact tracing apps can correctly assess the potential infection risk of a user with the virus if the following four important factors are considered:

\begin{enumerate}
    \item The distance between the infected person and the user;
    \item The length of time the infected person and the user occupy the same space;
    \item How many days prior to infection the infected person interacted with the user;
    \item Whether the user is likely to have touched a contaminated surface after interacting with the infected person.
\end{enumerate}

Developing corresponding rules based on the above factors helps establish the basic effectiveness of the developed contact tracing technologies. 

\subsubsection{Complete and accurate user data.} Digital contact tracing technologies are largely data-driven and the tracking mechanisms works based on the analysis on the collected user data in order to produce the close contacts should an encounter occurs with infected people. 

\subsubsection{Human and societal factors.} Contact tracing is a systematic process which goes beyond technologies themselves. Many human and societal factors exert strong impacts on the effectiveness of contact tracing such as the level of user adoption, government's proactiveness in disease control as well as local regulations and laws.

\subsection{Energy Efficiency}

As contact tracing is primarily carried out on smartphones, then the battery life of smartphones is a major technical factor that determines the availability of contact tracing when it is needed. The battery life of smartphones are impacted by a wide variety of factors, such as the processor, the battery size, the screen size, the screen brightness, the screen-on time, and the number and types of the running apps, etc. If we only focus on the energy efficiency of contact tracing apps, then it is largely determined by the protocols used in localization and data exchange. 

Different localization technologies, such as Bluetooth, GPS, cellphone tower networks, Wi-Fi and acoustic-ranging feature different levels of energy efficiency, where cellphone tower networks and Wi-Fi are comparatively more energy efficient, while GPS is more power hungry. 

From the perspective of protocols, battery consumption is highly related to the number of times data are exchanged with the server. For apps that utilize a centralized architecture, a fixed size of information is periodically retrieved from the server to obtain a new TempID. In contrast, there is no periodic data retrieval during the operational phase for the apps based on a decentralized architecture. Therefore, apps based on a decentralized architecture are usually more power efficient than those based on a centralized architecture. 

% \subsection{Interference}

% Contact-tracking applications, especially those that allow individuals to self-report that they are infected, must address the risk of false reporting by some individuals. In some cases, the false reporting may be bonafide - the person genuinely suspects that they are infected with COVID-19, but they have not been explicitly tested and are infected by a different virus. In other cases, a bad actor may report that they are infected with COVID-19 to create confusion. Storing sensitive information in an anonymized, redacted, and aggregated manner minimizes the risk of data tampering, but it does not eliminate the opportunity for human error or malicious intervention. One way to reduce fraud is to require confirmation of a diagnosis by a health care provider. However, creative teams may find other ways to prevent fraudulent disease reporting.

\subsection{Scalability}

After a contact tracing solution is being framed, developed and deployed, it is then important to understand and analyze how it behaves under certain computational parameters and assumptions. Given the typical large user base of contact tracing apps, it is technically challenging for the system to efficiently handle the large number of user visits and their tracing data generated. Apps that lack good scalability may be subject to frequent system crash which will seriously disrupt the normal operation of contact tracing.

Achieving a good scalability performance sets high requirements on the efficiency of contact tracing technologies in many aspects, such as the overall efficiency in data processing and analytic algorithms, data exchange and communication, and database query processing. In many cases, stress tests are necessary to assess the scalability of the system prior to its deployment to see whether it, as a whole, can operate reliably under various high workloads for a sustained duration.

\subsection{Compatibility}

% 设备系统兼容性
Smartphones are running on different types and versions of operating systems (OS), with Android and iOS being the two main ones. Because digital contact tracing was only started very recently in some countries, most tracing apps are developed for running on newer versions of OS to take advantages of the latest localization and security features offered by the operating system. For example, TraceTogether and CovidSafe both require iOS 10 or later. TraceTogether requires Android 5.1 or later, while CovidSafe can run on Android 6.0 or later. CovidSafe requires a newer OS version for security reasons and improved Bluetooth functionality. Given this, some problems may be caused due to the OS incompatibility issue, making some users unable to participate in the contact tracing because of the incompatibility between their smartphones and the contact tracing apps. 

Also, incompatibility may happen across different apps. Let's consider a case when a user who is using an app released for a particular geographic region travels to an area where another app has been deployed. Currently, there are no contact tracing apps that can directly talk to each other to share functions or data. In addition,due to the possible architectural differences discussed earlier, incompatibility between multiple apps may happen which may lead to conflicting ways the data are shared and used. Thus, users should be aware of the potential implications that may be caused by cross-app incompatibility in terms of privacy issues and other concerns.

%%%comment 1.3
\subsection{Social Issues}
Contact tracing technology is an effective solution to solve social problems such as in the widespread spread of COVID-19, but it may create social problems that may not be easy to be addressed. For example, social inequality will be caused by forcing citizens to use a certain tracing technology and the discriminatory measures on the traced movements of people. This will need to digital social exclusion and inequality. Social conflicts, even instability, may also be caused between governments and those people who believe in personal freedom around the mandated tracing measures. The extent in which contact tracing is utilised in our society also needs to be carefully measured. In the crisis of COVID-19, we will be more inclined to sacrifice the freedom as our response to the crisis. But when the situations of COVID-19 pandemic is in some countries, we may not have some appropriately relaxed contact tracing available accordingly. The current tracing apps rarely support different levels of tracing stringency that are adaptive varying pandemic situations.

Therefore, the development and implementation of contact tracing technologies should not be regarded as a social practice \cite{jt47}. In the long run, the design and implementation of contact tracing technologies should be based on common public values, privacy protection and democratic procedures, otherwise it may undermine public trust and national unity.

% individuals could be required to use contact tracing apps if they wished to participate in certain activities (e.g enter public/private spaces) or use public transport (Parker et al. 2020).These restrictions would make the app de facto compulsory if individuals are to remain functioning members of society, and would result in discrimination against those who either refuse to use it or do not have smartphones. 

\subsection{Cultural Issues}
Even though contact tracing has been generally perceived as an effective weapon against the pandemic of COVID-19, unfortunately, we have observed quite different effectiveness of contact tracing in curbing the spread of disease in different countries. Cross-cultural studies has supported that each specific culture has its own beliefs related to particular explanations for health and sickness. For the most part, East Asia and oceanian  countries are global leaders in preventing the spread of COVID-19 because of a vigilant public concerned for public safety and the good compliance with public safety measures, while countries in other continents have been (seriously) struggling with the pandemic. It appears that collectivism (or embeddedness) have become an important social factor that influences the overall effectiveness of COVID-19 control in terms of reducing its infection and fatality rates. This collectivism is based on a strong centralised authority leading a vigilant population that is concerned with the public safety of others as well as effective planning, communication, and enforcement of public safety measures \cite{Cultural} including contact tracing.

\subsection{Legal Issues}
In order to control COVID-19 spread, sharing one's personal and biometric information with a health professional or a research institute in contact tracing is not a violation of law. But if the data are accessed without proper authorization or misused, Remedial measures, and sometimes even legal actions, should be taken in a timely manner to safeguard the privacy and integrity of the data.

In addition, the government may impose mandatory public health measures based on the contact tracing results to limit the moments of some people who have been infected by the virus as well as their close contacts. However, in some countries, large numbers of people have voiced their opposition against this as infringements to their constitutionally protected freedom.

\subsection{Ethical Issues}
Contact tracing also pose some potential ethical challenges on the global scale, which are inevitably intervened with the aforementioned social, cultural and legal issues.

It is generally perceived that the use of tracking apps should be an individual choice on a voluntary basis. However, in some countries, individuals have been required to use contact tracing apps if they wish to participate in certain activities or move to certain locations. The pressing ethics issue emerges for compulsory adoption of contact tracing apps which may lead to the divide of social groups of people with the group of people whose freedoms and equity are compromised. 

Also, the transparency, oversight, and accountability also the important ethical issues need to be addressed for contact tracing. Transparency in contact tracing is critical to maintain its legitimacy and gain public confidence and trust. Citizens deserve clarity about many aspects of the implementation and the use of contact tracing apps. Also, who will oversee the implementation of contact tracing and ultimately take the responsibility for any major issues or failures of such a large-scale practice, should be clearly specified.

%Lastly，coercion. Apart from feasibility (not everyone owns or is able to use a smartphone), the coerced use of an app is at odds with a liberal democratic constitutional state. In a democracy, the autonomy of citizens is respected by allowing citizens to make their own decisions as much as possible. Voluntary use of the app is therefore a key condition.}

% \subsection{Equity}

% Contact Tracing apps shall be developed in collaboration with the privacy and security community, human rights and civil liberties organizations, government agencies, technology community, and public health professionals, including epidemiologists.

% \subsubsection{Efficiency in reducing the spread of infectious diseases}
% One of the advantages of smartphone-based contact tracing technology is its speed. When an infected person is detected, his/her contacts can be traced almost immediately. In a traditional contact investigation, it may take several days to obtain these prior contacts. Mobile contact tracing technology has shown tremendous impact on individual quarantines. First, its speed is absolutely critical. Secondly, it must be precise in detecting real contacts.

% %算法追踪精度
% \subsubsection{Algorithmic tracking accuracy}
% A common way of representing interactions between individuals is through the use of network diagrams. These network diagrams can be a very useful tool for understanding infection processes in human populations.

\section{challenges and future directions}

    Despite the unprecedented development of contact tracing technologies for COVID-19 witnessed in the last 12 months, there are still several major challenges that the tracing technologies are facing, which are elaborated as follows.
    
   \subsection{Difficulties in User Adoption} 
        
        The ultimate success of contact tracing is largely dependent on how many users are willing to participate in the tracing. Only when a significant percentage of users participating in contact tracing can the tracing become truly effective
        %%%comment 1.6
        \cite{jt25}. Conversely, user adoption has become a major obstacle for contact tracing. In some countries, contact tracing using digital means have turned out to be a failure. One example is Australians CovidSafe app, which was labeled by recently by Australian media as a costly failure. 

        Arguably, the primary reason for users' hesitation in participating in the tracing lies in the concerns around the privacy and security issues of the contact tracing technologies.
        %%%comment 1.6
        Some contact tracing systems openly publish data about patients diagnosed with COVID-19, trading off some of their privacy to enhance the privacy of individuals who are trying to determine if they have been exposed \cite{jt26}. Despite great efforts that have been taken by contact tracing system and app developers in enhancing the preservation of privacy information of users and strengthening system security against possible hacking and attacks which try to illegally access users' sensitive information, it is still a very challenging task to persuade users to adopt contact tracing technologies in many countries, particularly in the Western countries where the culture of individual privacy is deep rooted. We have witnessed some individual cases of mandatory adoption of contact tracing technologies in several countries under some special circumstances. For example, all the arriving passengers in South Korea and Singapore from overseas were once required, with no exception, to use their contact tracing app upon passengers' arrival. However, in a more border scale, we believe that user adoption has been, and will continue to be, the primary obstacle for contact tracing technologies.

       The technological solutions applied to deal with the privacy issue often create the challenges in balancing user privacy and data utility. While centralized storage of data can bring efficiency to governments and support research and other activities, it raises concerns that data may be stored or re-purposed for other uses. Thus, part of the equation to solve or mitigate the privacy concerns in contact tracking may be to avoid centralized data collection. However, a decentralized approach to data collection in contact tracing reduces the opportunity for health researchers and epidemiologists to easily access data for comprehensive studies of populations, leading to reduced data utility overall. For example, the UK government's proposal to adopt Apple/Google products has reduced public health surveillance and analysis. More importantly, reliance on Apple and Google for this technology has the clear potential to place more data mining and public health impact opportunities in the hands of the already dominant technology companies.
        
        %\subsection{Technical safety} 
        
        %Several technology-based solutions have been proposed for adoption as solutions/mechanisms for digital haptic tracking, some of which are based on GPS tracking, while others are based on Bluetooth token sharing. Despite these efforts, there are still issues with the underlying technology that should be addressed to narrow any strategic window of opportunity for malicious actors, snoopers, and surveillance states/governments. The Bluetooth-based contact tracking system can directly detect whether the user is coming in close proximity to the other party. The proximity can be approximated by the signal strength, although the signal strength can be reduced by obstacles such as walls. It can therefore be a more effective and accurate reflection of functional proximity in high-risk proximity environments such as buildings or public transportation. However, for Bluetooth-based applications to assess exposure risk, proximity exchange is inherently insufficient because, in addition to person-to-person interaction, COVID-19 can also be transmitted through common environments or surfaces of common contact. On the other hand, GPS is by its inherent nature insecure. In addition, some features that cannot be provided by GPS-based systems. One of the major problems is the spoofing attack, where a spoofer creates a false GPS signal to a particular receiver with the wrong time and location.
        %\item \textbf{Single tracking method}
        %\item \textbf{Pervesive tracking }
       
       \subsection{Incomplete and Low-quality Data}
        
        Technically, effective contact tracing relies largely on the availability of underlying complete and high-quality user data collected in the first place. However, due to various objective and subject reasons, the collected data are often incomplete and of a low quality. From the objective point of view, the localization technologies, due to their inherent limitations, may not be able to generate complete and high-quality localization data for the systems or apps to establish the accurate human movement trajectory and the list of close contacts of people. The quality of data may also be compromised in the communication process from the tracing devices to the server. The smartphones, as the most important means of physical devices for contact tracing, may not be properly working all the time, e.g., they may be running out of power when the tracing is performed. From the subjective perspective, users may fail to open the tracing devices for collecting the data. This may because that they are reluctant to use the tracing devices to share the data or they simply forget to turn it on. This leads to the totally missing or incomplete tracing data being collected. Whatever the underlying reasons they might be, incomplete and low-quality data will undoubtedly make accurate and effective contact tracing very difficult, if it is not entirely impossible.
        
        \subsection{Lack of Scenario-specific Tracing Technologies}
        
        The current digital contact tracing technologies are all very generic in the overall configuration and setup.
        %%%comment3.19
        They are applied for contact tracing without scenario-specific customization and there are few technologies which are developed particularly for specific tracing scenarios from the very beginning. However, we understand that the tracing requirements for different scenarios  can be (quite) different in terms of, for example, the spatial location (indoor or outdoor), temporal information (weekdays vs. weekends), population density and mobility, etc. Such as, contact tracing  in indoor commercial and industrial spaces needs ultra-precise Wi-Fi-based local positioning systems \cite{jt29}. Such scenarios-specific factors will have impacts on the corresponding optimal specification of the localization technology, data collection and storage approaches, the values of some key tracing parameters as well as notification and response strategies. And it is problematic to use contact tracing technologies based on mobile devices alone in prison inmates, poverty-stricken neighborhoods, etc. where the number of smartphone is very little \cite{jt27}, and then we'll end up losing a whole bunch of people and not reaching the goal of 60-80 percent of the population using digital contact tracing to add value \cite{jt30}.
        
        %%修改前
        %The current digital contact tracing technologies are all very generic in the overall configuration and setup. They are applied for contact tracing without scenario-specific customization. Hence, using them directly in specific scenarios typically leads to subpar tracing performance. There are few technologies which are developed particularly for specific tracing scenarios from the very beginning. We understand that the tracing requirements for different scenarios (e.g., schools, hospitals or residential areas) can be (quite) different in terms of, for example, the spatial location (indoor or outdoor), temporal information (weekdays vs. weekends), population density and mobility, etc. Such scenarios-specific factors will have impacts on the corresponding optimal specification of the localization technology, data collection and storage approaches, the values of some key tracing parameters as well as notification and response strategies. 

        \subsection{Lack of Global Standards and Availability}
        %%%comment 1.6+3.20
        Apple and Google collaborated on a common contact-tracing platform -- the (Google/Apple) Exposure Notification (GAEN) system. This protocol is accessible to public-health agencies wishing to use it for their own apps through an application-programming interface (API) called the 'Exposure Notification API', which will allow these apps to log and receive data. There are many countries had received access to the protocol and adopted it, but the apps are not at all standardized across the EU \cite{jt28}. Several countries, such as France, for example, have pursued to collect contact data on central government servers, not on decentralized servers, in order to maintain records of personal information that can be used to assist in investigating cases. Governments in different countries tend to support locally developed digital technologies for contact tracing due to the reasons of fast rollout and removal of obstacles posed by the local laws and regulations. These have created a technological barrier for compatibility of the technologies used in different countries, so when people are traveling across regions and countries, they have no choice but to adopt different technologies. This also creates isolated data islands and makes efficient sharing of contact tracing data across different countries difficult.
        %%修改前
        % Governments in various countries and regions tend to support locally developed digital technologies for contact tracing due to the reasons of fast rollout and removal of obstacles posed by the local laws and regulations. Consequently, there haven't been international standards established yet for unified contact tracing technologies and protocols. We haven't seen any major initiatives towards this goal led by any countries or international organizations either.
        % This has created a technological barrier for compatibility of the technologies used in different countries, so when people are traveling across regions and countries, they have no choice but to adopt different technologies. This also creates isolated data islands and makes efficient sharing of contact tracing data across different countries difficult.
        \par In addition, digital contact tracing technologies at present are adopted and implemented in a very small number of countries in the world, overwhelming majority of which are in developed countries. This may be due to a higher percentage of population that use smartphones, better ICT as well as public health infrastructures, and more funding resources, etc in developed countries compared to developing countries. Because of the unavailability of the contact tracing in most countries in the world, the chasm has been created in tracing contacts for international travelers when they are traveling from or to the countries where contact tracing are lacking. 

    \subsection{Future Directions}
    Looking ahead, we envisage that the contact tracing technologies will carry some new characteristics in the future, which can be concisely described using the following several terms: privacy, intelligence, pervasiveness, scenario customization and globalization. 
    
    \begin{itemize}
    \item \textbf{Privacy.} Contact tracing technologies in the future will become increasingly safer and provides a better protection of users sensitive information. Data collected and stored in the servers and mobile devices, either in the centralized or distributed architecture, will be better protected against malicious attacks to avoid the disclosure of sensitive data to adversaries. Novel architectures may also emerge to offer better preservation of sensitive users data than the existing architectures do. Relevant laws, policies or regulations will be developed by different countries or international organizations to better regulate the use and sharing of user data collected in contact tracing. All these developments will be conducive to the higher level of acceptance and adoption among the public for contact tracing technologies, which are of great importance to reduce the spread of virus;
    \item \textbf{Intelligence.} Contact tracing technologies, thanks to the fast development of various intelligent technologies in the areas of machine learning, big data analytics and data visualization, etc, will become more intelligent and are capable of providing a rich set of analytical features than just the tracking function that most of the existing tracing systems and apps deliver. Such intelligence in contact tracing technologies will allow them to better tap the high values of the large amount of user data collected;
    \item \textbf{Pervasiveness.} Contact tracing technologies will become more pervasive and ubiquitous to supply more complete data in support of high-quality contact tracing. This will be empowered by various technological advancements including, but not limited to, the  wider usage of smartphones in the general public, more power efficient tracing devices and more accessible and reliable communication infrastructures;
    \item \textbf{Scenario customization.} To achieve the best contact tracing performance, contact tracing technologies need to undergo scenario-specific customization by incorporating scenario-specific background facts and knowledge, enabling them be applied in a wide variety of application scenarios, such as schools, hospitals, factories and companies, etc., in a more effective and efficient manner;
    \item \textbf{Globalization.} Contact tracing systems or apps will become broader in terms of their geographic coverage, pushing for the gradual (or rapid) globalization of the technologies. Along with this development, standardization of the involved technologies will be naturally entailed to create unified international standards in a bid to facilitate cross-border contact tracing among different countries. 
    
    \end{itemize}

Looking ahead, we are confident that contact tracing, whatever form it will take, will continue to play an indispensable role in our human being's continuous and unrelenting combat against infectious diseases into the future.

\section{Conclusion}
In this paper, we present a comprehensive survey on contact tracing, covering both the traditional and the recent digital contact tracing technologies. The fundamental approach and models of contact tracing are covered and, as the focus of this paper, we survey the latest digital contact tracing technologies, detailing a variety of contact tracing mobile apps currently being used in many different countries, together with the associated localization technologies. The important issues and related protocols are also discussed. At the end of the survey, we reflect on the current challenges contact tracing technologies are facing and highlight several future trends projected for technological developments of contact tracing in the future.

Given the rampant COVID-19 pandemic we are currently experiencing, we believe that this survey work is relevant and significant. We hope that this survey can help readers better understand the history, the current status and the future trends of contact tracing technologies, and provide a good reference to researchers to research and develop future-generation conduct tracing technologies to help curb the spread of deadly virus. 

\bibliographystyle{IEEEtran}
\bibliography{CT_main}

\end{document}